\documentclass[a4paper,fleqn,usenatbib]{mnras} 

\usepackage{latexsym}
\usepackage{amssymb}
\usepackage{amsfonts}
\usepackage{amsmath}
\usepackage{bm}
\usepackage{graphicx}
\usepackage{subfigure}
\usepackage{times}
\usepackage{units}
\usepackage{hyperref}
\usepackage{multirow}
\usepackage{comment}
\usepackage{ulem}
\usepackage{hyperref}
\usepackage{graphicx}
\usepackage{acronym}
\usepackage{xcolor}
\usepackage{booktabs}

\usepackage{pifont} 

\usepackage{enumitem}
\setlist[itemize]{noitemsep, topsep=0pt}

\usepackage{multirow}
\usepackage{morefloats}
\usepackage{mathrsfs}
\usepackage{latexsym}
\usepackage{dcolumn}
\usepackage{lipsum}
\usepackage{mathtools}
\usepackage{cuted}
\renewcommand{\cite}[1]{\citep{#1}}
\renewcommand\eqref[1]{(\ref{#1})}

\newcommand{\GW}{GW170817}
\newcommand{\AT}{AT2017gfo} 
\newcommand{\GRB}{GRB170817A}
\newcommand{\swind}{spiral-wave wind}

\newcommand{\mr}{mass ratio} 
 
\newcommand{\bnc}{bounce} 

\newcommand{\rproc}{$r$-process}
\newcommand{\nuc}{nucleosynthesis}

\newcommand{\surname}[1]{\text{#1}}

\newcommand{\ie}{\textit{i.e.}}
\newcommand{\eg}{\textit{e.g.}}

\newcommand{\be}{\begin{equation}}
    \newcommand{\ee}{\end{equation}}
\newcommand{\bea}{\begin{eqnarray}}
    \newcommand{\eea}{\end{eqnarray}}
\newcommand{\bel}{\begin{align}}
    \newcommand{\eel}{\end{align}}

\newcommand{\pba}{\texttt{PyBlastAfterglow}}

\def\Msun{{\rm M_{\odot}}}
\def\GMc2{{\rm G M_{\odot} c^{-2}}}

\def\O{\mathcal{O}}

\def\ccm{\,\text{cm}^{-3}}
\def\size{FWHM$_{x}$}

\newacro{MRI}{magnetorotational instability}

\newacro{BH}{black hole}
\newacro{BNS}{binary neutron star}
\newacro{EM}{electromagnetic}
\newacro{EOS}{equation of state}
\newacroplural{EOS}[EOSs]{equations of state}
\newacro{GR}{general relativity}
\newacro{HD}{hydrodynamics}
\newacro{MHD}{magnetohydrodynamics}
\newacro{PIC}{particle-in-cell}
\newacro{BPL}{broken power law}
\newacro{SNR}{supernova remnant}
\newacro{BM}{Blandford \& McKee}
\newacro{ST}{Taylor-von Neumann-Sedov}
\newacro{FBOT}{fast blue optical transient}
\newacro{SSA}{synchrotron self-absorption}
\newacro{SED}{spectral energy distribution}
\newacro{GRHD}{general-relativistic hydrodynamics}
\newacro{GRMHD}{general-relativistic magnetohydrodynamics}
\newacro{GW}{gravitational wave}
\newacro{BZ}{Blandford–Znajek}
\newacroplural{GW}[GWs]{gravitational waves}
\newacro{LES}{large-eddy simulation}
\newacroplural{LES}[LES]{large-eddy simulations}
\newacro{GRLES}{general-relativistic large-eddy simulation}
\newacro{NR}{numerical relativity}
\newacro{NS}{neutron star}
\newacroplural{NS}[NSs]{neutron stars}
\newacro{GRB}{gamma-ray burst}
\newacroplural{GRB}[GRBs]{gamma-ray burst}
\newacro{kN}{kilonova}
\newacroplural{kN}[kNe]{kilonovae}
\newacro{SGRB}{short \ac{GRB}}
\newacro{ISM}{interstellar medium}
\newacro{MNS}{massive neutron star}
\newacro{NSBH}{neutron star-black hole}
\newacro{NSE}{nuclear statistical equilibrium}
\newacro{SN}{supernova}
\newacroplural{SN}[SNe]{supernovae}
\newacro{CCSN}{core-collapse supernova}
\newacroplural{CCSN}[CCSNe]{Core-collapse supernovae}
\newacro{MP}{metal-poor}
\newacro{UFG}{ultra-faint dwarf galaxy}
\newacroplural{UFG}[UFGs]{ultra-faint dwarf galaxies}
\newacro{NRN}{nuclear reaction network}
\newacroplural{NRN}[NRNs]{nuclear reaction networks}
\newacro{RR}{reaction rate}
\newacroplural{RR}[RRs]{reaction rates}
\newacro{ODE}{ordinary differential equation}
\newacroplural{ODE}[ODEs]{ordinary differential equations}
\newacro{IR}{infrared}
\newacro{NIR}{near-infrared}
\newacro{FIR}{far infrared}
\newacro{LC}{light curve}
\newacroplural{LC}[LCs]{light curves}
\newacro{UV}{ultraviolet}
\newacro{LTE}{local thermodynamic equilibrium}
\newacro{TOV}{Tolman-Oppenheimer-Volkoff}
\newacro{RMS}{root mean square}
\newacro{MM}{multi-messenger}
\newacro{LK}{light curve}
\newacro{CBM}{circumburst medium}
\newacro{LF}{Lorentz factor}
\newacro{EFE}{Einstein’s field equations}
\newacro{ADM}{Arnowitt, Deser and Misner}
\newacro{IVP}{initial value problem}
\newacro{RHS}{right hand side}
\newacro{RMF}{relativistic mean-field}
\newacro{PDE}{partial differential equation}
\newacroplural{PDE}[PDEs]{partial differential equations}
\newacro{EM}{electromagnetic}
\newacro{HMNS}{hyper-massive neutron star}
\newacro{SMNS}{supra-massive neutron star}
\newacro{EATS}{equal time arrival surface}
\newacroplural{EATS}[EATSs]{equal time arrival surfaces}
\newacro{JWST}{James Webb Space Telescope}
\newacro{VLA}{Very Large Array}
\newacro{SKA}{Square Kilometre Array}
\newacro{DSA}{Diffusive Shock Acceleration}
\newacro{PN}{post-Newtonian}
\newacro{EOB}{effective-one-body}
\newacro{PC}{prompt collapse}
\newacro{LIGO}{Laser Interferometer Gravitational-Wave Observatory}
\newacro{MF}{magentic field}
\newacroplural{MF}[MFs]{magnetic fields}
\newacro{AGN}{active galactic nucleus}
\newacroplural{AGN}[AGNs]{active galactic nuclei}
\newacro{LOS}{line of sight}
\newacro{DE}{dynamical ejecta}
\newacro{SWW}{\swind{}}

\newacro{BW}{blast wave}
\newacroplural{BW}[BWs]{blast waves}

\newacro{VLBI}{very-long-baseline interferometry}
\newacro{ISS}{interstellar scintillation}
\newacro{mas}{milliarcsecond}
\newacro{FWHM}{full width at half maximum}

\renewcommand{\cite}[1]{\citep{#1}}
\defcitealias{Margalit:2021kuf}{\scshape MP21} 
\defcitealias{Nedora:2021eoj}{\scshape N21} 
\defcitealias{Nedora:2022kjv}{\scshape N22a} 

\title[BNS afterglow skymaps]{
    Synthetic radio images of structured GRB and kilonova afterglows
}

\author[V.\ Nedora et al.]{
    Vsevolod \surname{Nedora}$^{1,2}$, 
    Tim \surname{Dietrich}$^{1,2}$,
    Masaru \surname{Shibata}$^{1,3}$
    \\
    ${}^1$Max Planck Institute for Gravitational Physics (Albert Einstein Institute), Am M{\"u}hlenberg 1, Potsdam 14476, Germany\\
    ${}^2$Institute for Physics and Astronomy, University of Potsdam, Potsdam 14476, Germany\\
    ${}^3$Center for Gravitational Physics and Quantum Information, Yukawa Institute for Theoretical Physics, Kyoto University, Kyoto, 606-8502, Japan
}

\date{
    Accepted XXX. Received YYY; in original form ZZZ
}

\begin{document}
\label{firstpage}

\maketitle

\date{\today}

\begin{abstract}
     In this paper, we compute and analyze synthetic radio images 
     of gamma-ray bursts and kilonova afterglows. For modeling the former, we consider  
     \GRB{}-inspired set of parameters, while for the latter, we employ ejecta 
     profiles from numerical-relativity simulations. 
     We find that the kilonova afterglow sky map has a doughnut-like structure at early times that becomes more ring-like at late times.
     This is caused by the fact that the synchrotron 
     emission from electrons following Maxwellian 
     distribution function dominates the early, beamed, emission while emissions from electrons following power-law distribution is important at late times.
     For an on-axis observer, the image flux centroid moves on the image plane initially away from the observer. 
     The image sizes, we find, are the largest for equal mass merger simulations with  
     the soft equation of state. 
     The presence of a kilonova afterglow affects the properties inferred from 
     the source sky map even if the gamma-ray burst afterglow dominates the total flux density.  
     The main effect is the reduction of the mean apparent velocity of the source, 
     and an increase in the source size. 
     Thus, neglecting the presence of the kilonova afterglow may lead to systematic 
     errors in the inference of gamma-ray burst properties from the sky map observations. 
     Notably, at the observing angle inferred for \GRB{} the presence of 
     kilonova afterglow would 
     affect the sky map properties only at very late times $t\gtrsim1500\,$days.
\end{abstract}


\section{Introduction}
\label{sec:intro}

Radio observations have always played an important role in 
\ac{GRB} studies. Besides complementing the broadband spectrum 
analysis, they allow for direct and indirect 
measurements of the source geometry and dynamics. 
Specifically, just by using observations of total flux 
density around the \ac{LC} peak, it is very challenging to 
constrain the observing angle, \ie, the angle between the 
\ac{GRB} jet axis and the observer \ac{LOS}. 
This results in degeneracy among the model parameters 
\cite{Nakar:2020pyd}. Observations of the shift of the radio 
image centroid allow us to break this degeneracy. 
However, \ac{GRB} jets at cosmological distances are less then a 
parsec in size and thus their imaging is complicated even with 
the most sensitive \ac{VLBI} facilities. Examples of a 
successful imaging include \ac{GRB}030329 and \GRB{}. 

\ac{GRB}030329 was imaged via global \ac{VLBI}, that reached 
sub-\ac{mas} resolution. The image size, approximated with \ac{FWHM} 
(assuming a circular Gaussian model for the image) was 
$0.07\,$\ac{mas} and $0.17\,$\ac{mas} at $23\,$ and 
$83$ days, respectively \cite{Taylor:2004wd}. Multiple observations 
at different epochs yielded an average expansion of $3-5\,c$. 
This superluminal motion hinted at a relativistic expansion of 
the \ac{GRB} jet. 
This source was also imaged $217\,$days \cite{Taylor:2004ru} and 
$806\,$days \cite{Pihlstrom:2007zz} after the original trigger.
Combined analysis of the radio images and broadband data 
yielded estimates on the jet parameters and its 
lateral spreading as well as on the angle beween jet axis and the 
\ac{LOS} \cite{Granot:2004qf,Pihlstrom:2007zz,Mesler:2012,Mesler:2013fza}.

Another example of a successful jet imaging is \GRB{} 
\citep{Savchenko:2017ffs,Alexander:2017aly,Troja:2017nqp,Monitor:2017mdv,Nynka:2018vup,Hajela:2019mjy}, 
a short \ac{GRB} detected by the space observatories 
Fermi \citep{TheFermi-LAT:2015kwa} and INTEGRAL \citep{Winkler:2011} 
and localized to the S$0$ galaxy NGC$4993$. 
\GRB{} was an electromagnetic counterpart to the \ac{GW} even 
\GW~\citep{TheLIGOScientific:2017qsa,Abbott:2018wiz,LIGOScientific:2018mvr}. 
This \ac{GRB} was dimmer than other events of its class and 
followed by an afterglow with a prolonged rising part.
The most widely accepted explanation for this is that \GRB{} was a 
structured jet observed off-axis 
\citep[\eg][]{Fong:2017ekk,Troja:2017nqp,Margutti:2018xqd,Lamb:2017ych,Lamb:2018ohw,Ryan:2019fhz,Alexander:2018dcl,Mooley:2018dlz,Ghirlanda:2018uyx}.
This interpretation is in contrast to the commonly considered uniform jet structure, 
also called ``top-hat'' \cite{Rhoads:1997ps,Panaitescu:1998zf,Sari:1999mr,Kumar:2000gj,Moderski:1999ct,Granot:2001cw,Granot:2002za,Ramirez-Ruiz:2004cvd,Ramirez-Ruiz:2004gvs}, 
where energy and momenta do not depend on the angle 
(outside the jet opening angle). 
This explanation was in part derived from the analysis of radio images 
at  $75\,$ and $230\,$ days after the burst by the Karl G. Jansky \ac{VLA} 
and the Robert C. Byrd Green Bank Telescope \cite{Mooley:2018dlz}. 
The observations showed that the position of the flux centroid has 
changed between two observational epochs, with the mean apparent velocity 
along the plane of the sky $\beta_{\rm app}=4.1 \pm 0.5$. 
The source, however, remains unresolved.
That gave a possible upper limit on the source 
size of $1\,$mas and $10\,$\ac{mas}  
in the direction perpendicular and parallel to the motion 
respectively \cite{Mooley:2018dlz}. 
The high compactness of the source was further supported 
by the observed quick turnover around the peak of the radio \acp{LC} 
and a steep decline $F_{\nu}\propto t_{\rm obs}^{-2}$ after $200\,$days 
\cite{Mooley:2018clx}.
%
Notably, the superluminal motion was also observed in the optical band
\cite{Mooley:2022uqa}. 
\citet{Ghirlanda:2018uyx} also obtained a radio image at $207\,$days 
confirming the previous findings. Together with the analysis of multi-wavelength 
\acp{LC}, the information obtained from radio images allowed to 
confirm that \GRB{} was produced by a narrow, core-dominated jet rather 
than by a wide, quasi-isotropic ejecta \cite{Hotokezaka:2018gmo,Gill:2018kcw}. 
A comparison with \ac{GRB}030329, where no proper motion was 
observed, only the expansion speed, indicates a difference in 
source geometry.

A sizable fraction of \acp{GRB} occurs further off-axis than \GRB{}. 
For them, the prompt $\gamma$-ray emission, as well as early afterglow 
may not be seen as they would be beamed away from the observer's \ac{LOS}. 
At later times, however, as the jet decelerates and spreads laterally, the 
afterglow should become visible. Such afterglow is referred as 
``orphan afterglow'' \cite{Rhoads:1997ps}. No such afterglow 
has been found so far despite extensive search campaigns in 
X-ray \cite{Woods:1999vx,Nakar:2002un}, optical \cite{Dalal:2001ym,Totani:2002ay,Nakar:2002ph,Rhoads:2003ya,Rau:2006}, 
and radio \cite{Perna:1998ny,Levinson:2002aw,Gal-Yam:2005gbc,Soderberg:2005vp,Bietenholz:2013hha} 
(see also \citet{Huang:2020pxr}).

In addition to \ac{GRB} and its afterglow, \GW{} was 
accompanied by a quasi-thermal electromagnetic counterpart, \ac{kN}  \AT~\citep{Arcavi:2017xiz,Coulter:2017wya,Drout:2017ijr,Evans:2017mmy,Hallinan:2017woc,Kasliwal:2017ngb,Nicholl:2017ahq,Smartt:2017fuw,Soares-santos:2017lru,Tanvir:2017pws,Troja:2017nqp,Mooley:2018dlz,Ruan:2017bha,Lyman:2018qjg}. 
The ejecta responsible for the \ac{kN} was enriched with heavy elements, 
lanthinides and actinidies, produced via \rproc{} \nuc{} \citep{Lattimer:1974slx,Li:1998bw,Kulkarni:2005jw,Rosswog:2005su,Metzger:2010,Roberts:2011,Kasen:2013xka,Tanaka:2013ana}.
The angular and velocity distributions of these ejecta are quite  
challenging to infer due to the complex atomic properties of 
these heavy elements. Nevertheless, at least two ejecta components 
to account for the observed \acp{LC} were needed: a lanthanide-poor 
(for the early blue signal) and a lanthanide-rich (for the late red signal) one \cite{Cowperthwaite:2017dyu,Villar:2017wcc,Tanvir:2017pws,Tanaka:2017qxj,Perego:2017wtu,Kawaguchi:2018ptg,Coughlin:2018fis}. 
A fit of \AT{} \acp{LC} to a semi-analytical two-components 
spherical \ac{kN} model yielded blue (red) components of mass
$2.5\times10^{-2}M_{\odot}$ ($5.0\times10^{-2}M_{\odot}$) 
and velocity $0.27$c ($0.15$c) \citep{Cowperthwaite:2017dyu,Villar:2017wcc}. 
The estimated ejecta mass and velocity could be significantly modified if anisotropic effects are taken into account~\cite{Kawaguchi:2018ptg}.

\Ac{NR} simulations of \ac{BNS} mergers predict that 
mass ejection can be triggered by different mechanisms acting 
on different timescales (see 
\citet{Metzger:2019zeh,Shibata:2019wef,Radice:2020ddv,Bernuzzi:2020tgt} 
for reviews on various aspects of the problem). 
Specifically, dynamical ejecta of mass $\O(10^{-4}-10^{-2})\,\Msun$ 
can be launched during mergers at average velocities of $0.1-0.3\,$c, \eg~
\cite{Rosswog:1998hy,Rosswog:2005su,Hotokezaka:2013iia,Bauswein:2013yna,Wanajo:2014wha,Sekiguchi:2015dma,Radice:2016dwd,Sekiguchi:2016bjd,Vincent:2019kor,Zappa:2022rpd,Fujibayashi:2022ftg}.
After the merger, quasi-steady state winds were shown to emerge 
from a post-merger disk  \cite{Dessart:2008zd,Fernandez:2014bra,Perego:2014fma,Just:2014fka,Kasen:2014toa,Metzger:2014ila,Martin:2015hxa,Wu:2016pnw,Siegel:2017nub,Fujibayashi:2017puw,Fahlman:2018llv,Metzger:2018uni,Fernandez:2018kax,Miller:2019dpt,Fujibayashi:2020qda,Nedora:2020pak,Nedora:2019jhl}.

\ac{NR} simulations also show that a small fraction of dynamical ejecta 
$({\sim}(10^{-6}-10^{-5})\,\Msun)$ has velocity exceeding ${\simeq}0.6\,c$, 
\citep{Hotokezaka:2013b,Metzger:2014yda,Hotokezaka:2018gmo,Radice:2018pdn,Radice:2018ghv,Nedora:2021eoj,Fujibayashi:2022ftg}. 
Such a fast ejecta are capable of producing 
bright non-thermal late-time afterglow-like emission, with \ac{SED} peaking 
in radio band \citep[\eg][]{Nakar:2011cw,Piran:2012wd,Hotokezaka:2015eja,Radice:2018pdn,Hotokezaka:2018gmo,Kathirgamaraju:2018mac,Desai:2018rbc,Nathanail:2020hkx,Hajela:2021faz,Nakar:2019fza}. 
The mechanisms behind the fast tail of the ejecta is not yet clear. 
Possible options include shocks launched at core \bnc{} 
\citep{Hotokezaka:2013b,Radice:2018pdn} and shocks generated 
at the collisional interface between \acp{NS} \cite{Bauswein:2013yna}.

Notably, despite a large amount of \ac{BNS} \ac{NR} simulations 
there is no robust relationship between the binary parameters 
\acp{NS} and \ac{NS} \ac{EOS} and the properties of the ejected matter. 
And while there exist fitting formulae of various complexity to 
the properties of the bulk of the ejecta, \eg, mass and velocity 
\cite{Dietrich:2016fpt,Radice:2018pdn,Kruger:2020gig,Nedora:2020qtd,Dietrich:2020efo}, 
even such formulae for the fast ejecta tail are currently absent. 
Thus, we are limited to employing published dynamical ejecta profiles 
from \ac{NR} simulations. 

In \citet{Nedora:2021eoj} (hereafter \citetalias{Nedora:2021eoj}) we 
showed how a \ac{kN} afterglow emission from the fast tail of the dynamical 
ejecta may contribute to the radio \acp{LC} of the \GRB{}, employing 
\ac{NR}-informed ejecta profiles \cite{Perego:2019adq,Nedora:2019jhl,Nedora:2020pak,Bernuzzi:2020tgt}. 
In \citet{Nedora:2022kjv} (hereafter \citetalias{Nedora:2022kjv}) we modified 
the afterglow model by including an additional electron population that 
assumes Maxwellian distribution in energy behind the \ac{kN} \ac{BW} shock.
We showed that the radio flux from these ``thermal electrons'' 
can be higher than the radio flux from commonly considered 
``power-law'' electrons at early times and if ejecta is sufficiently fast. 
It is thus natural to investigate, whether the emission from 
thermal electrons affects the radio image of the source. 
As in \citetalias{Nedora:2022kjv} we consider a \GRB{}-inspired 
\ac{GRB} afterglow model of a Gaussian jet, seen off-axis, while 
for \ac{kN} afterglow we consider \ac{NR}-informed ejecta profiles 
extracted from \ac{NR} simulations with various \ac{NS} \acp{EOS} and 
system \mr{}s.

The paper is organized as follows. 
In Sec.~\ref{sec:method}, we recall the main assumptions and methods 
used to calculation the observed \ac{GRB}- and \ac{kN}- afterglow 
emission as well as how to compute the sky map. 
In Sec.~\ref{sec:result:kn}, we present and discuss the \ac{kN} 
afterglow sky maps, focusing on the overall properties, \eg, image size 
and the flux centroid position and their evolution.
In Sec.~\ref{sec:result:kn_grb}, we consider both \ac{GRB} and \ac{kN} 
afterglow and discuss how the properties of the former change when the 
later is included in the modeling. 
In Sec.~\ref{sec:result:interaction}, we briefly remark on how the 
\ac{GRB} plus \ac{kN} sky map changes if the \ac{ISM} in front of the 
\ac{kN} ejecta has been pre-accelerated and partially removed by a 
passage of the laterally spreading \ac{GRB} ejecta. 
Finally, in Sec.~\ref{sec:conclusion}, we provide the discussion and conclusion.

\begin{figure*}
    \centering 
    \includegraphics[width=0.49\textwidth]{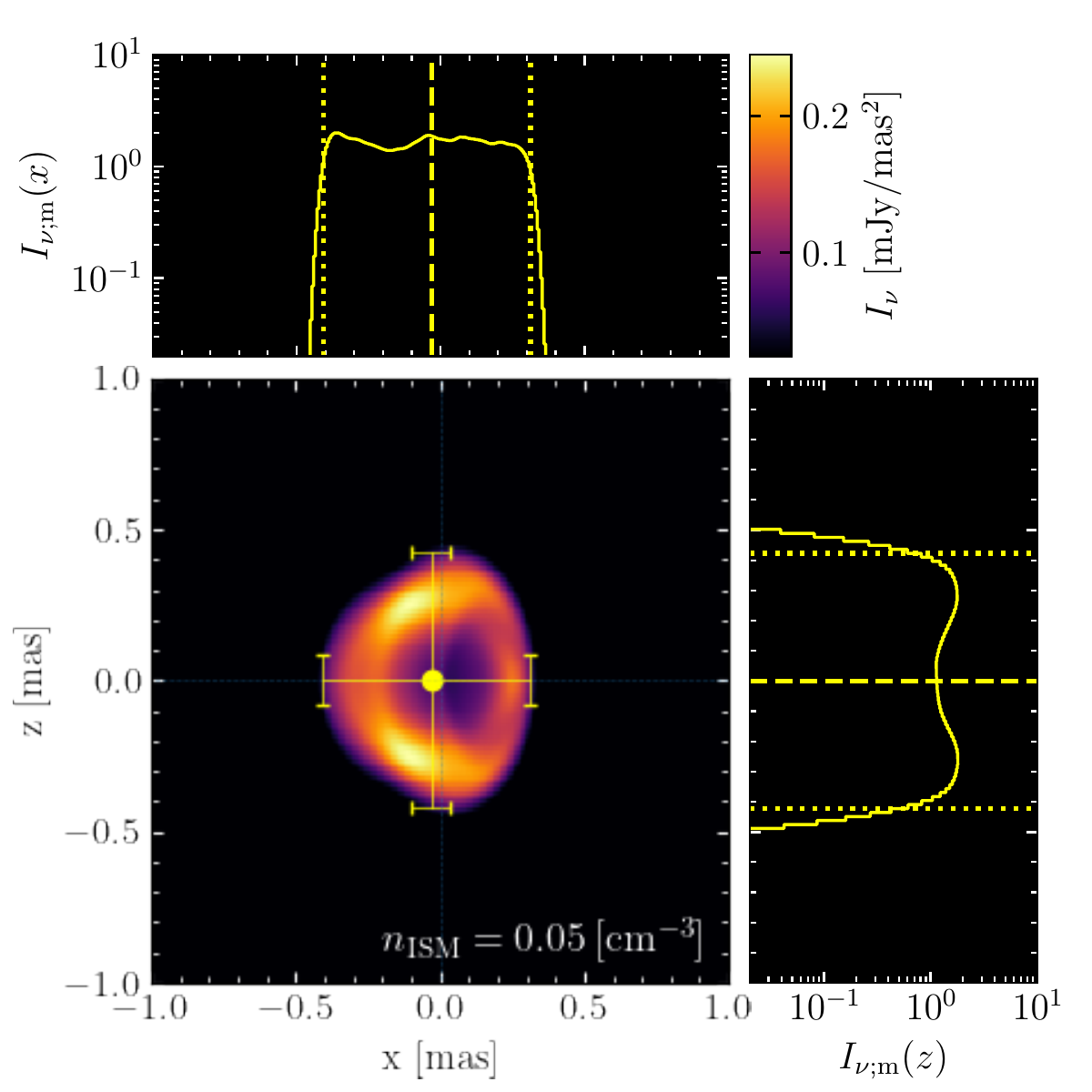}
    \includegraphics[width=0.49\textwidth]{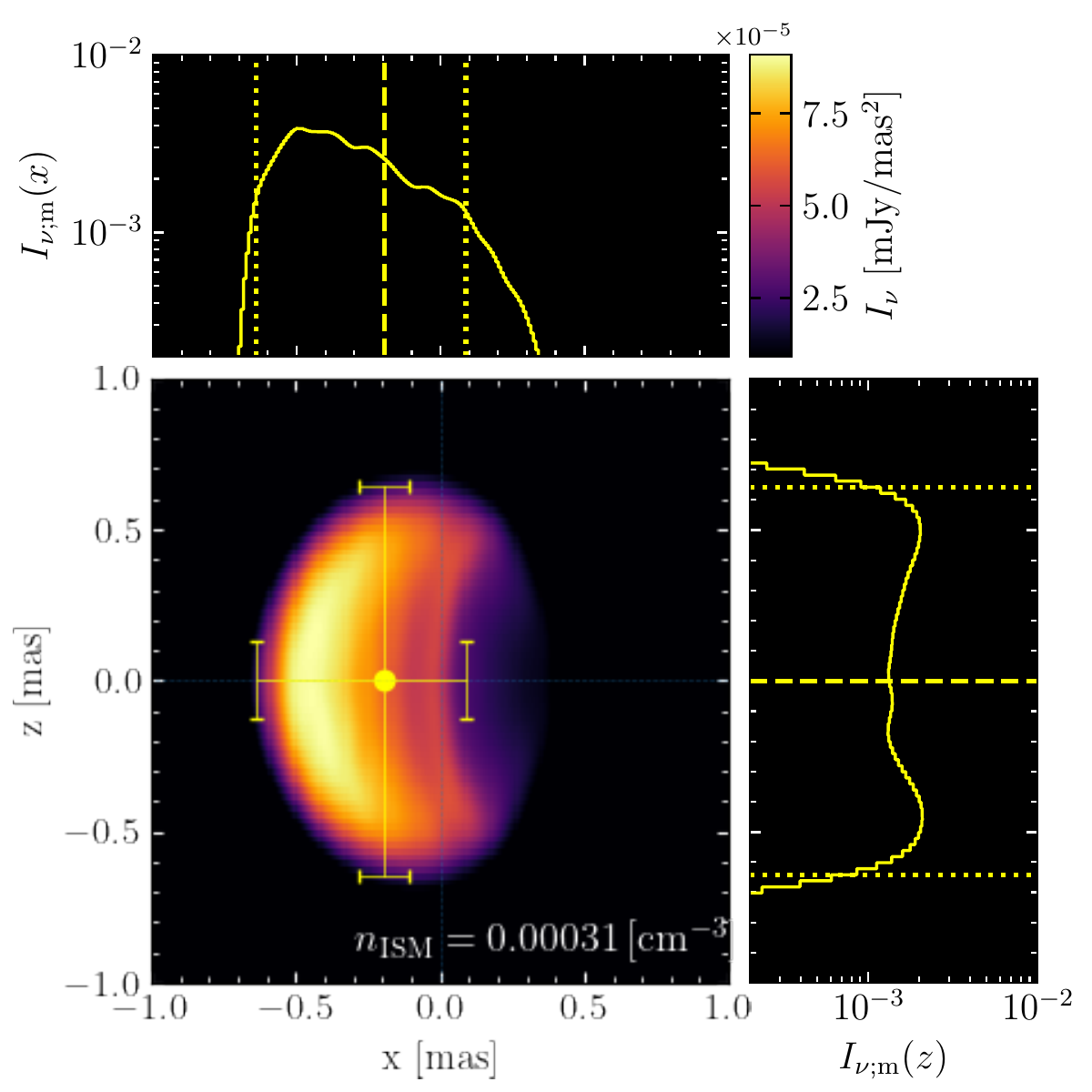}
    \includegraphics[width=0.49\textwidth]{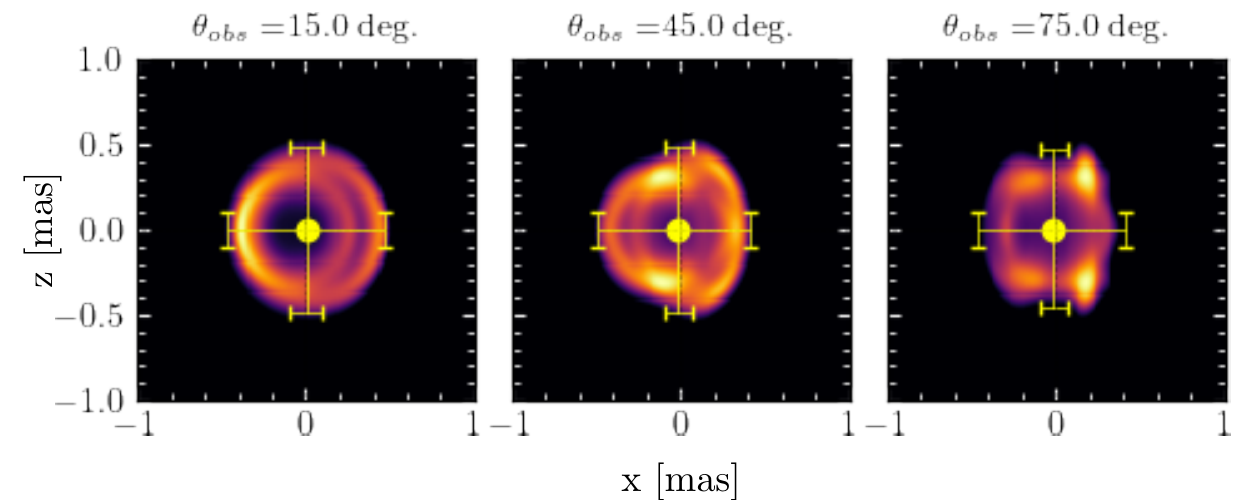}
    \includegraphics[width=0.49\textwidth]{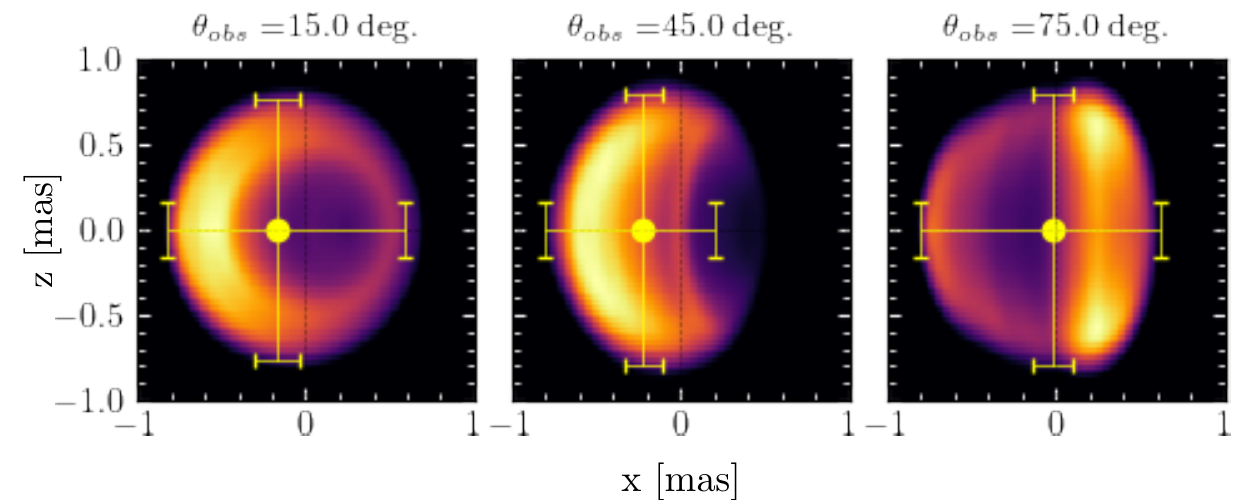}
    \caption{
        \textit{Top figure}:
        sky map for a \ac{BNS} merger simulation with BLh \ac{EOS} and 
        $q=1.00$ computed at $120\,$ days after the merger at 
        $\nu=1\,$GHz and observed at $\theta_{\rm obs}=45\,$deg. 
        Left and right columns of plots corresponds to different \ac{ISM} densities, 
        $n_{\rm ISM}=0.05\,\ccm$ on the left and $n_{\rm ISM}=0.00031\,\ccm$ on the right. 
        In each plot column, the top and top-right subplots display the 
        $X$ and $Z$ averaged brightness distributions 
        respectively. Dotted lines mark \ac{FWHM} and 
        dashed lines mark the location of the flux 
        centroid of the image. \ac{FWHM} and 
        the location of the flux centroid are also shown on 
        the main panel of the figure as error bars and the 
        circular marker respectively. 
        Thin gray dotted lines indicate the $X$ and $Z$ axis. 
        Notably, we are plotting $I_{\nu}/I_{\nu;\,\rm max}\in (0.1,1)$ range of the normalized 
        specific intensity in order to resolve the image structure 
        more clearly. 
        \textit{Bottom panel}: same sky map but viewed at 
        three difference angles, $\theta_{\rm obs}$. 
    } 
    \label{fig:results:skymap_example}
\end{figure*}

\section{Methods} \label{sec:method}

In order to compute the \ac{GRB} and \ac{kN} afterglows, we employ the 
semi-analytic code \pba{} discussed in \citetalias{Nedora:2022kjv} and 
\citetalias{Nedora:2021eoj}. In the model, both ejecta 
types are discretized into velocity and angular elements, for each 
of which the equations of \ac{BW} evolution are solved independently, 
and the synchrotron radiation is computed, accounting for relativistic 
and time-of-arrival effects. The effect of the pre-processing of 
\ac{ISM} medium by a passing \ac{GRB} \ac{BW} is considered in 
Sec.~\ref{sec:result:interaction}, otherwise this effect is not included, 
and \ac{kN} \acp{BW} evolved independently from the \ac{GRB} \acp{BW}. 
For \ac{kN} afterglow, both thermal and non-thermal electron populations 
are considered, while for \ac{GRB} afterglow only the latter is employed 
in the model. 

The sky maps are computed using the spherical coordinate system discussed 
in Sec.~$2$ in \citetalias{Nedora:2022kjv} (figure~1). For both ejecta types  
axial symmetry is assumed. Then, each elemental \ac{BW} has 
radial coordinate $R_{ij}$, and angular coordinates $\theta_{i}$ and 
$\phi_{ij}$, where the single index of $\theta_{i}$ reflects the 
axial symmetry. 
The coordinate vector of the elemental \ac{BW} is given by
$\vec{v}_{ij} = R_{ij}\big( \sin{(\theta_{i})}\cos{(\phi_{ij})}\vec{x},\, \sin{(\theta_{i})}\sin{(\phi_{ij})}\vec{y},\, \cos{(\theta_{i})}\vec{z} \big)$. 
The cosine of the angle between the \ac{LOS} and $\vec{v}_{ij}$ reads,  
\begin{equation}  \label{eq:method:mu_obs}
    \mu_{ij} = 
    \sin{(\theta_{i})}\sin{(\phi_{ij})}\sin(\theta_{\rm obs}) +  
    \cos{(\theta_{i})}\cos(\theta_{\rm obs}) \, .
\end{equation}
The image plane, $xz$ is perpendicular to the \ac{LOS} 
of the observer. We chose the basis with which the 
principal jet moves in the positive $\tilde{x}$-direction.  
The basis vectors then $\tilde{\vec{x}}_{ij}=\sin(\theta_{\rm obs})\vec{z}_{ij}-\cos(\theta_{\rm obs})\vec{x}_{ij}$, 
$\tilde{\vec{y}}_{ij}=\vec{x}_{ij}$
of the plane as in \citet{Fernandez:2021xce} 
and the coordinates of the $ij$ \ac{BW} 
on the image plane (for the principle jet) are given by 
\begin{equation}\label{eq:method:xccoord}
\begin{aligned}
    \tilde{x}_{ij} & = -R_{ij}[\cos(\theta_{\rm obs})\sin(\theta_i)\sin(\phi_{ij}) \\
           & + \sin(\theta_{\rm obs})\cos(\theta_{i}))], \\
    \tilde{z}_{ij} & = R_{ij}\sin(\theta_{i})\cos(\phi_{ij})\, . 
\end{aligned}
\end{equation}
In the following, we omit the use of tildas for simplicity.

\begin{figure*}
    \centering 
    \includegraphics[width=1.0\textwidth]{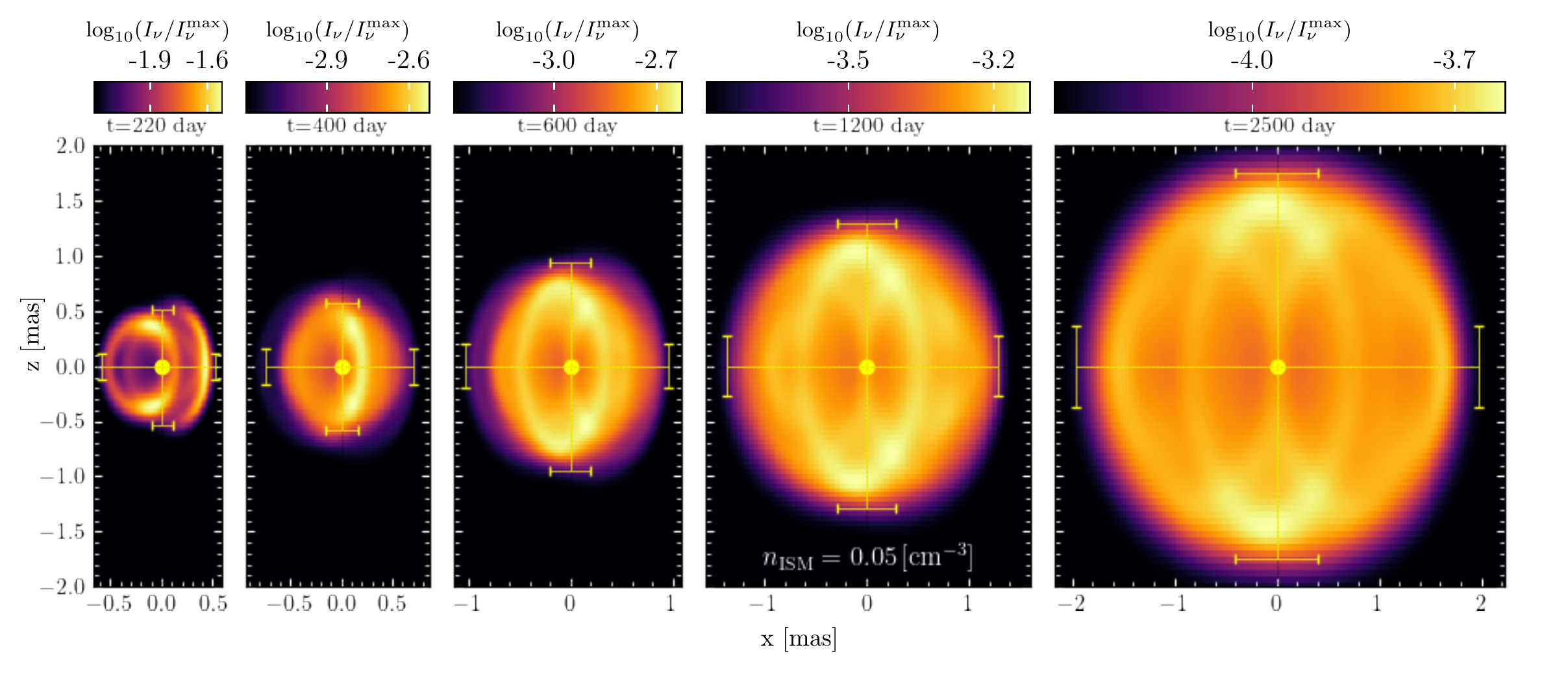}
    \caption{
        Evolution of the sky map for the \ac{BNS} merger simulation 
        with BLh \ac{EOS} and $q=1.00$, observed at $\theta_{\rm obs}=45\,$deg, $\nu_{\rm obs}=1\,$GHz. The \ac{ISM} density is 
        $n_{\rm ISM}=0.05\,\ccm$.
        As in Fig.~\ref{fig:results:skymap_example}, 
        the marker and the error bar indicate the location of the flux centroid 
        and the \ac{FWHM} of the image, while gray dotted lines mark the axis. 
    } 
    \label{fig:results:kn_skymap_example_evolution}
\end{figure*}

\begin{figure*}
    \centering 
    \includegraphics[width=1.0\textwidth]{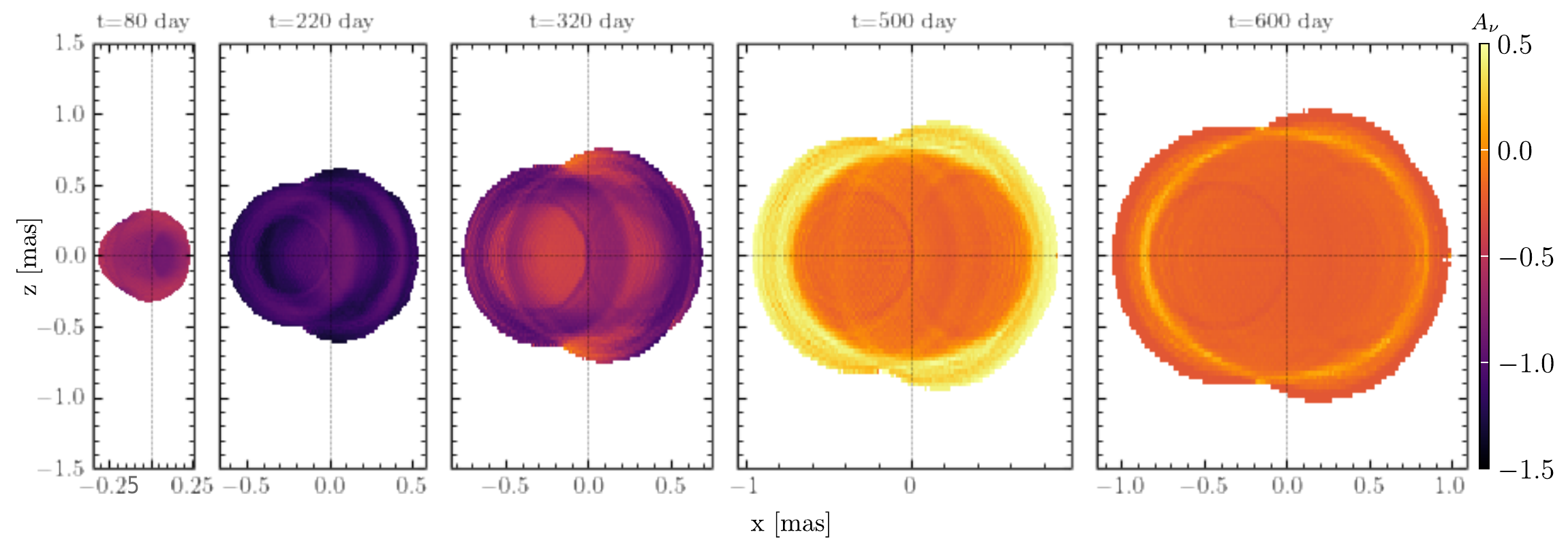}
    \caption{
        Evolution of the sky map spectral index for the \ac{BNS} merger simulation 
        with BLh \ac{EOS} and $q=1.00$, observed at $\theta_{\rm obs}=45\,$deg. 
        and at $\nu=1\,$GHz. Here $n_{\rm ISM}=0.05\,\ccm$.
        Thin dotted lines mark the axis. 
        For clarity, we did not apply the Gaussian smoothing kernel to this image. 
    } 
    \label{fig:results:kn_skymap_spectral_example_evolution}
\end{figure*}

In order to characterize sky maps we consider the following 
main quantities. Specifically, following \citet{Zrake:2018eml,Fernandez:2021xce} 
we compute the surface brightness-weighted center of the image, 
image centroid, defined as 
\begin{equation}
    x_c = \frac{1}{\int I_{\nu} dx dz}\int x I_{\nu} dx dz,
\end{equation}
where $I_{\nu}$ is computed via Eq.~(37) in \citetalias{Nedora:2022kjv}. 
We also compute the $X$ and $Z$-averaged brightness distributions 
\begin{equation}
    \begin{aligned}
    I_{\nu; \rm m}(x) &= \frac{1}{\Delta z} \int I_{\nu}(x,z) dz\, , \\
    I_{\nu; \rm m}(z) &= \frac{1}{\Delta x} \int I_{\nu}(x,z) dx\, .
\end{aligned}
\end{equation}
As the available ejecta profiles are limited in the angular 
resolution, which severely limits the accuracy of the sky map analysis,
we ``rebin'' the angular ejecta distribution histograms. 
To do this rebinning we assume a uniform distribution within 
each bin \cite{Knoll:2000fj}.

\section{Results} \label{sec:result}

For an extended source with uniform \ac{LF} $\Gamma$, 
the maximal apparent velocity $\beta_{\rm app} < \Gamma$, 
while the image size increases with $\Gamma$ 
\cite{Boutelier:2011}. A spherically symmetric source 
that expands isotropically, would appear as a ring 
expanding with $\Gamma$ with no motion in the image centroid. 

Due to non-trivial ejecta angular and velocity structure
a \ac{kN} sky map shape, size and structure have a complex dependency 
on the observer time $t_{\rm obs}$ and angle $\theta_{\rm obs}$. 
Moreover, if both, thermal and non-thermal 
electron populations are present behind the shock, there is a 
non-trivial dependency on the microphysical parameters and \ac{ISM} density. 
It is beyond the scope of this work to study all possible combinations 
of free parameters. Instead, we focus on several 
representative cases.

Specifically, we fix the source to be located at luminosity distance, 
$D_L=41.3\,$Mpc with redshift $Z=0.0099$. The micropthysics parameters 
are the following. 
Fractions of the shock energy that goes into electron acceleration and magnetic field amplification are $\epsilon_e=0.1$, $\epsilon_b=0.001$, $\epsilon_t=1$.
The slope of the power-law electron distribution is $p=2.05$. 
Unless stated otherwise, the observational frequency is $1\,$GHz., and the observer angle is $45\,$deg.
We focus on two $n_{\rm ISM}$: the fiducial value $n_{\rm ISM} = 0.05\,\ccm$ and the value inferred for \GRB{}, $n_{\rm ISM} = 0.00031\,\ccm$.

\begin{figure*}
    \centering 
    \includegraphics[width=0.49\textwidth]{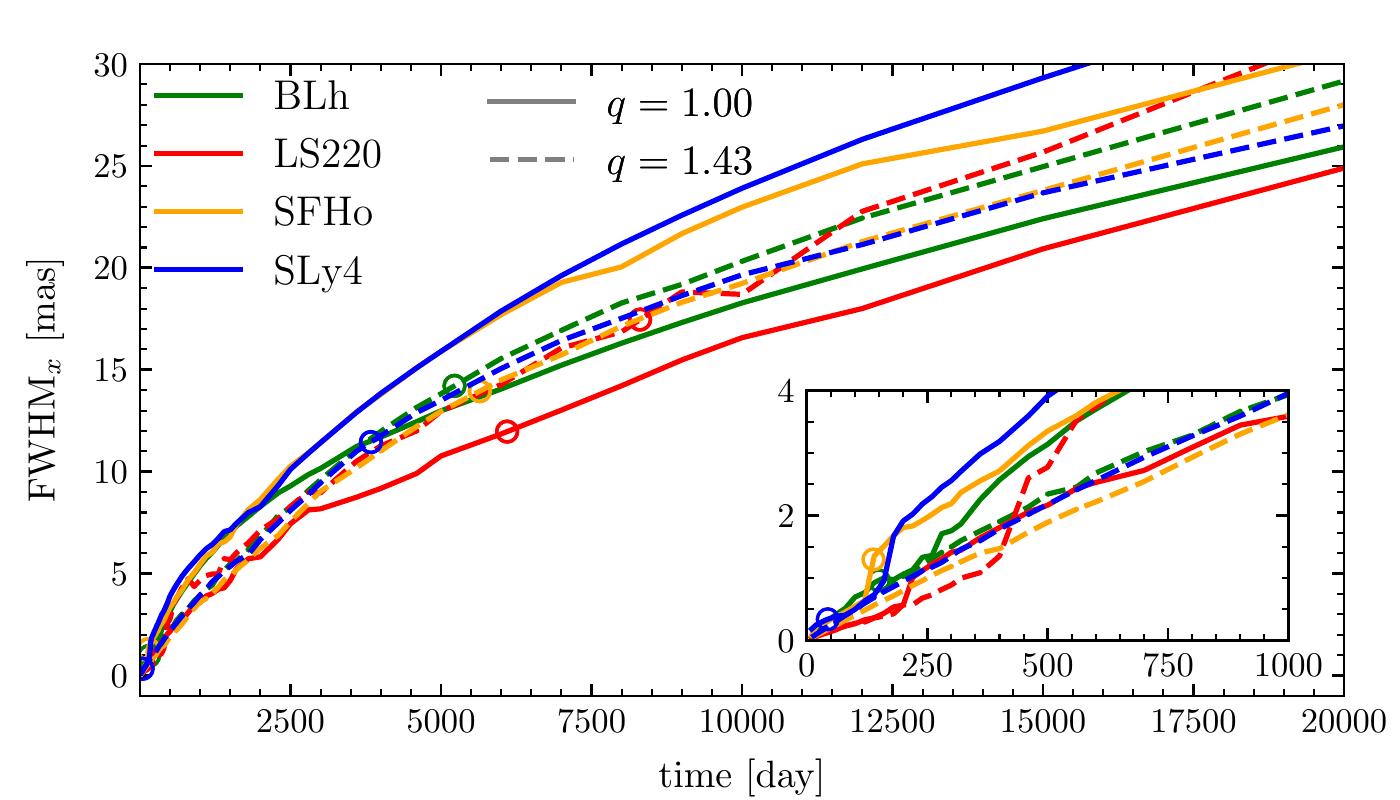}
    \includegraphics[width=0.49\textwidth]{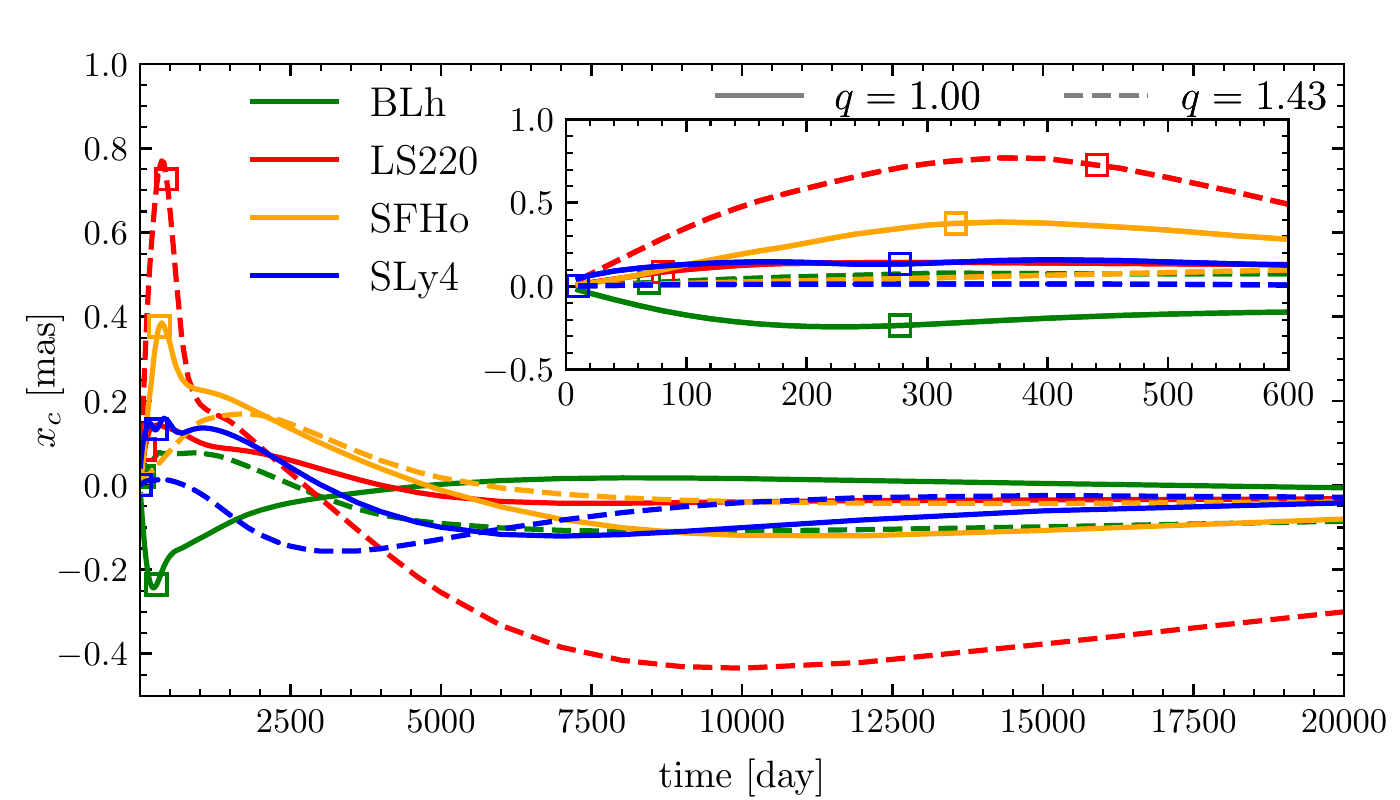}
    \includegraphics[width=0.49\textwidth]{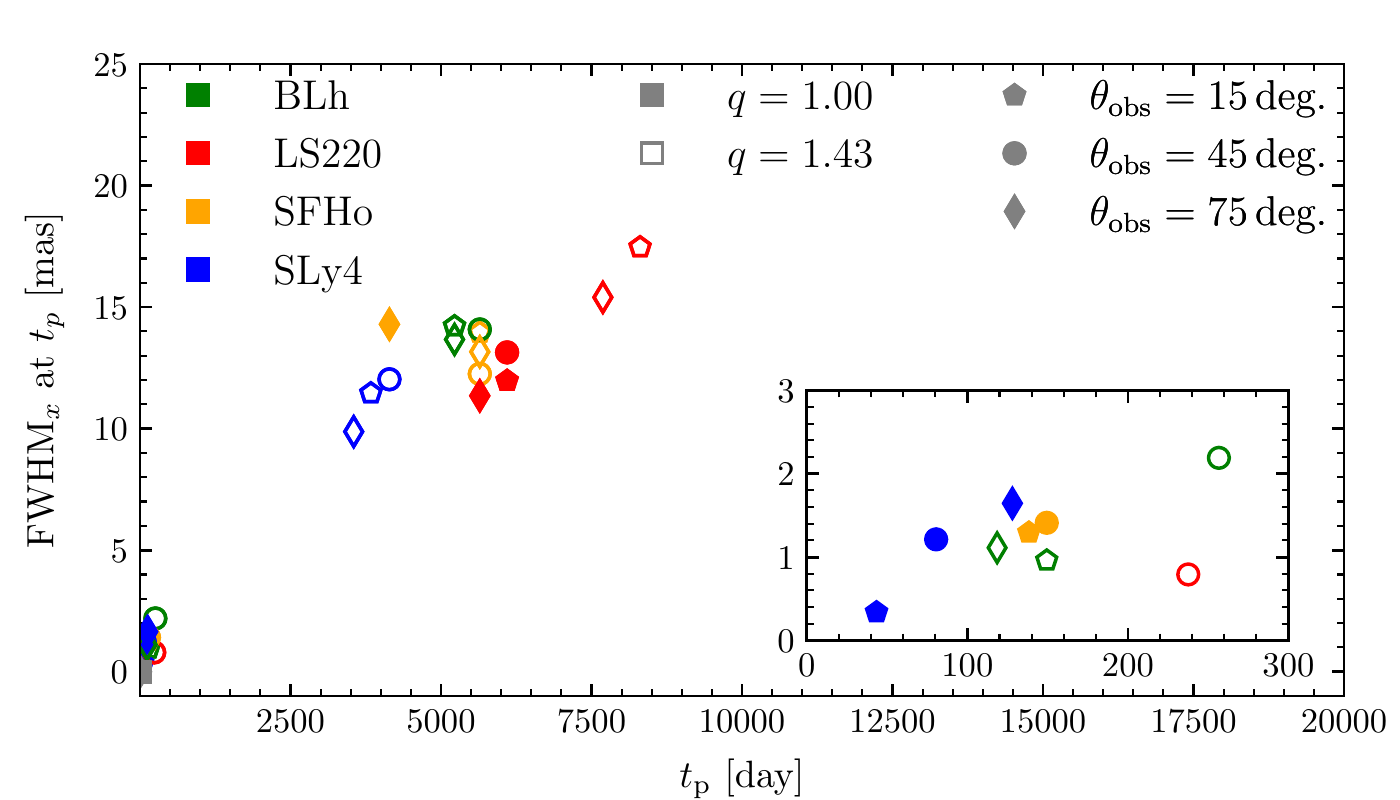}
    \includegraphics[width=0.49\textwidth]{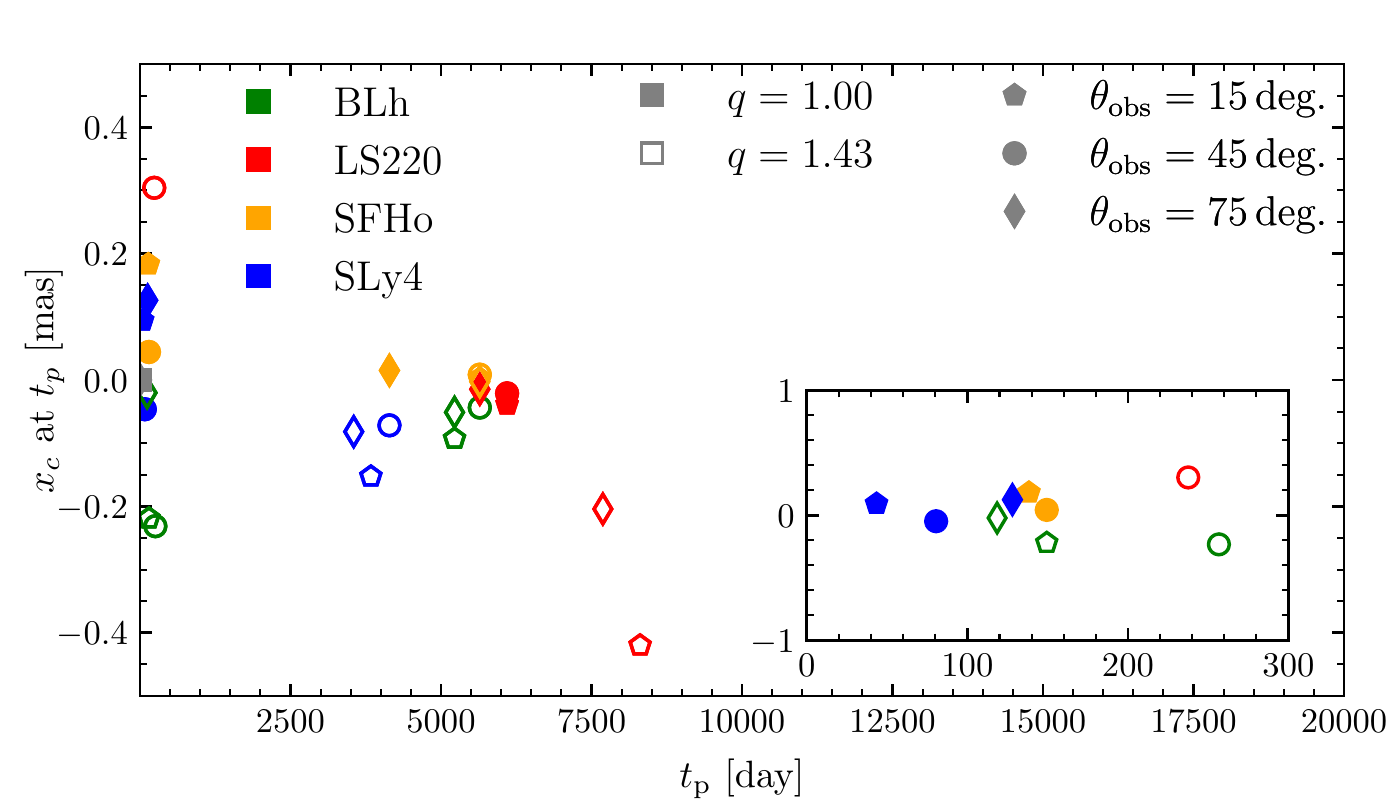}
    \caption{
        Time evolution of \ac{kN} afterglow sky map properties. 
        \textit{Top left panel} shows the evolution of the image size, \size{} 
        Circular markers indicate the image size at the \ac{LC} peak. 
        \textit{Top right panel} shows the evolution of the flux centroid, $x_c$ position. 
        Square markers indicate the minimum of the \ac{LC} spectral index $A_{\nu}$. 
        \textit{Bottom left panel} and \textit{Bottom right panel} display the 
        image size and the position of the flux centroid at the time of the 
        \ac{LC} peak respectively for three values of the observational angle. 
        In each panel there is a subpanel, enlarging an early-time part of the 
        plot. 
        Here $n_{\rm ISM} = 0.00031\,\ccm$. 
    } 
    \label{fig:results:kn_skymaps_width_evol_all_sims}
\end{figure*}

\subsection{Kilonova afterglow sky maps} \label{sec:result:kn}

We begin by considering the \ac{BNS} merger simulation 
with BLh \ac{EOS} and $q=1.00$.
The sky map for $\theta_{\rm obs}=45\,$deg, $\nu_{\rm obs}=1\,$GHz 
and $t_{\rm obs}=120\,$days after merger 
is shown in Fig.~\ref{fig:results:skymap_example}.

At $t_{\rm obs}=120\,$days the \ac{kN} afterglow at $1\,$GHz for this 
\ac{BNS} merger model is dominated by the emission from 
thermal electron population behind shocks. 
The fast tail of the dynamical ejecta in this simulation 
is predominantly equatorial, confined to ${\gtrsim}60\,$deg 
(see Fig.~3 in \citetalias{Nedora:2021eoj}) with mass-averaged 
half-\ac{RMS} angle $\theta_{\rm RMS}\simeq70\,$deg. 
As $\theta_{\rm RMS} > \theta_{\rm obs}$, the synthetic image 
resembles a wheel with the brightest parts offset from the 
center into the negative half of the $x$ axis, \ie, $x_c < 0$. 
(See $\theta_{\rm obs}=15\,$deg. and $\theta_{\rm obs}=45\,$deg. 
subpanels in Fig.~\ref{fig:results:skymap_example}). 
An observer with $\theta_{\rm obs} \gtrsim \theta_{\rm RMS}$ would, 
however, be able to see the beamed emission from the fast ejecta tail 
(bright spots at $x\simeq0\,$\ac{mas}, $z\simeq\pm0.3\,$\ac{mas} on 
$\theta_{\rm obs}=75\,$deg. sub-panels of 
Fig.~\ref{fig:results:skymap_example}). 
Correspondingly, the image flux centroid lies near 
$x_c \simeq 0$ and the brightest part of the image laying in $x>0$ plane.

As the \ac{kN} \acp{BW} propagates through the \ac{ISM}, the 
size of a sky map increases in both $x$ and $z$ directions. 
Due to the axial symmetry of the ejecta properties, 
$\theta_{\rm obs}$ and relativistic effects primarily affect 
the \size{} and $x_c$. The example of a sky map evolution is 
shown in Fig.~\ref{fig:results:kn_skymap_example_evolution}. 
Deceleration of \ac{kN} \acp{BW} reduces the contribution 
from thermal electron population to the observed flux. 
Additionally, relativistic effects become increasingly 
less important. Consequently, the image becomes more 
spherically symmetric and centered around $x_c=z_c=0$. 
Specifically for this simulation, at $t_{\rm obs}=600\,$days 
after the merger the emission from equatorial 
and polar \acp{BW} becomes comparable with each other and 
thereafter the sky map resembles a circle with two 
bright spots near the image's outer boundaries on the $x=0$ axis.
These spots mark the geometrically overlapping 
emitting areas and reflect the equatorial nature of the ejecta 
fast tail. 
Notably, the presence of thermal electrons that we assume in 
our model does not affect this qualitative picture, as the 
emissivity from both thermal and non-thermal electron populations 
depend on the shock velocity albeit to a different degree 
\citep[\eg][]{Ozel:2000,Margalit:2021kuf}.

The presence of two electron populations behind \ac{BW} 
shocks, however, implies a spectral evolution of the emission in every 
pixel of the sky map. We define a sky map spectral index as 
$A_{\nu} = d\log_{10}(I_{\nu})/d\log_{10}(\nu)$ and show its 
evolution in Fig.~\ref{fig:results:kn_skymap_spectral_example_evolution}
for $\theta_{\rm obs}=45\,$deg. and $\nu=1\,$GHz. 
At early times, most of the sky map displays relatively 
low $A_{\nu}\simeq-1.25$, indicative of the emission from thermal 
electron populations (figure~3 in \citetalias{Nedora:2022kjv}). 
As the \acp{BW} decelerate and emission 
from the thermal electron population subsides. At the point where 
the spectrum transitions, the spectral index reaches a minimum. 
After that, the spectral index 
rises as the sky map becomes increasingly dominated by emission 
from the non-thermal electron population. 
At very late times the spectral map becomes uniform, as the emission from 
power-law electrons with fixed distribution slope $p$ dominates in every 
pixel. 
If resolved in observations, such evolution of the spectral 
sky map would allow a detailed study of the ejecta velocity 
and angular distribution, besides constraining the physics of 
particle acceleration at mildly relativistic shocks.

It is interesting to examine the evolution of the key sky map 
properties, image size \size{}, and the position of the flux 
centroid, $x_c$, at very low \ac{ISM} density, that was generally 
inferred for \GRB{}. 
In Fig.~\ref{fig:results:kn_skymaps_width_evol_all_sims}, we show 
the evolution of the \size{} and $x_c$ as well as these 
values at the peak time $t_p$ of the respective \acp{LC}. 
The sky map size at a given epoch is primarily determined by the 
energy budget of the ejecta. Simulations with $q=1$ and soft 
\ac{EOS}, \eg, SFHo and SLy4 \acp{EOS} display larger image sizes 
throughout the evolution. 
On the other hand, equal mass simulations with stiffer \ac{EOS}, 
such as BLh and LS220 \acp{EOS} demonstrate smaller image sizes, 
More asymmetric binaries display in general intermediate image sizes. 

At the time of the \ac{LC} peak, the image size depends 
on whether the emission from thermal or non-thermal electron 
population dominates the observed flux. 
If former is true, $t_p$ is generally small, 
$t_p < 500\,$ days for our simulations and assumed 
$n_{\rm ISM}=0.00031\,\ccm$, and the image size case does not 
exceed $4\,$\ac{mas}. 
Notably, at higher $n_{\rm ISM}$ $t_p$ is shorter and 
thus, the \size{} is smaller. 
Simulations with $q=1.00$ and soft (SLy4 and SFHo) \acp{EOS} 
are examples of that. 
If the emission from power-law electrons dominates the observed 
flux at the time of the \ac{LC} peak, the image size is significantly 
larger, ${\simeq}15-20\,$\ac{mas}. Importantly, $t_p$ depends also 
on the observer angle $\theta_{\rm obs}$ due to relativistic beaming of 
the early-time emission from thermal electrons. 
For example, a simulation with a sufficiently spherically symmetric 
distribution of the fast tail, simulation with SLy4 \ac{EOS} and 
$q=1.00$ display an early $t_p < 500\,$days at all three observing 
angles considered.

A characteristic feature of the changing dominant contributor 
(\eg, electron population) to the observed emission is seen here 
as a sharp increase in the evolution of image size 
(sub-panel in the top left panel in 
Fig.~\ref{fig:results:kn_skymaps_width_evol_all_sims}). 
This rapid increase in \size{} occurs when the emission from fast 
\acp{BW}, dominating the observed flux at first, subsides and 
less beamed, more isotropic emission from non-thermal electrons 
becomes equally important.

As discussed before, the evolution of the image flux 
centroid position, $X_c$, besides the ejecta energy budget, 
depends strongly on the observational angle. 
At $\theta_{\rm obs}=45\,$deg. for \ac{BNS} merger models with 
sufficiently fast and equatorial fast tail, $x_c$ is 
negative at an early time (\eg, for simulation 
with BLh \ac{EOS} and $q=1.00$). For simulations with 
$\theta_{\rm RMS} < \theta_{\rm obs}$, $x_c$ moves into 
the positive half of $x$ axis at the beginning, as is the 
the case for the equal mass simulations with SFHo, SLy4 and LS220 
\acp{EOS}. 
The time evolution of the $x_c$ in most cases exhibits an 
extremum after which $x_c \rightarrow 0$. We find that 
the time of the extremum corresponds to the time where the 
spectral index evolution of the \ac{LC} reaches minimum 
(see figure~$4$ in \citetalias{Nedora:2022kjv} for the \ac{LC} 
spectral index evolution). 
In the top right panel of Fig.~\ref{fig:results:kn_skymaps_width_evol_all_sims} 
this point is shown with square marker.

At the time of the \ac{LC} peak the position of the 
flux centroid is generally determined by whether the thermal 
or non-thermal electrons dominate the observed flux. 
This in turn depends on $\theta_{\rm obs}$. 
In the former case $|x_c|$ tends to be larger, 
reaching $|x_c| \leq 0.5\,$\ac{mas}, as bright beamed emission 
from thermal electrons in fast \acp{BW} makes the image 
very asymmetric. Consequently, if the \ac{LC} peaks at 
late times, $|x_c|$ is closer to zero for most models.

\subsection{kN and GRB skymaps} \label{sec:result:kn_grb}

One of the key observables of \GRB{} that confirmed the jetted 
nature of the outflow and allowed for a more 
precise estimate of the inclination angle $\theta_{\rm obs}$, 
was the motion of the \ac{GRB} flux centroid \cite{Mooley:2018dlz}. 
Here we investigate, how the presence of the \ac{kN} afterglow 
affects the \ac{GRB} afterglow sky map $x_c$ and \size{}  
assuming that these two ejecta types do not interact. 
We briefly remark on this interaction 
in Sec.~\ref{sec:result:interaction}.

For modeling \ac{GRB} afterglows, we consider the same parameters 
as in \citetalias{Nedora:2022kjv}, motivated by the analysis 
of \GRB{} \citep[\eg][]{Hajela:2019mjy,Fernandez:2021xce}, 
varying only the observer angle, $\theta_{\rm obs}$ and 
the \ac{ISM} density $n_{\rm ISM}$. 
Specifically, we set the jet half-opening angle 
$\theta_{\rm w} = 15\,$deg. and core half-opening angle 
$\theta_{\rm c} = 4.9\,$deg. 
The isotropic equivalent energy is $E_{\rm iso}=10^{52}\,$ergs, 
and the initial \ac{LF} of the core is $\Gamma_{\rm c} = 300$. 
The microphysical parameters are set as:  
$\epsilon_e=0.05$, $\epsilon_B=0.0045$, and $p=2.16$. 
Luminosity distance to the source is set to $D_{\rm L} = 41.3\,$Mpc. 
Unless stated otherwise, we consider $\theta_{\rm obs}=45\,$deg., and  
$n_{\rm ISM} = 0.00031\,\ccm$, as fiducial values.

\begin{figure}
    \centering 
    \includegraphics[width=0.49\textwidth]{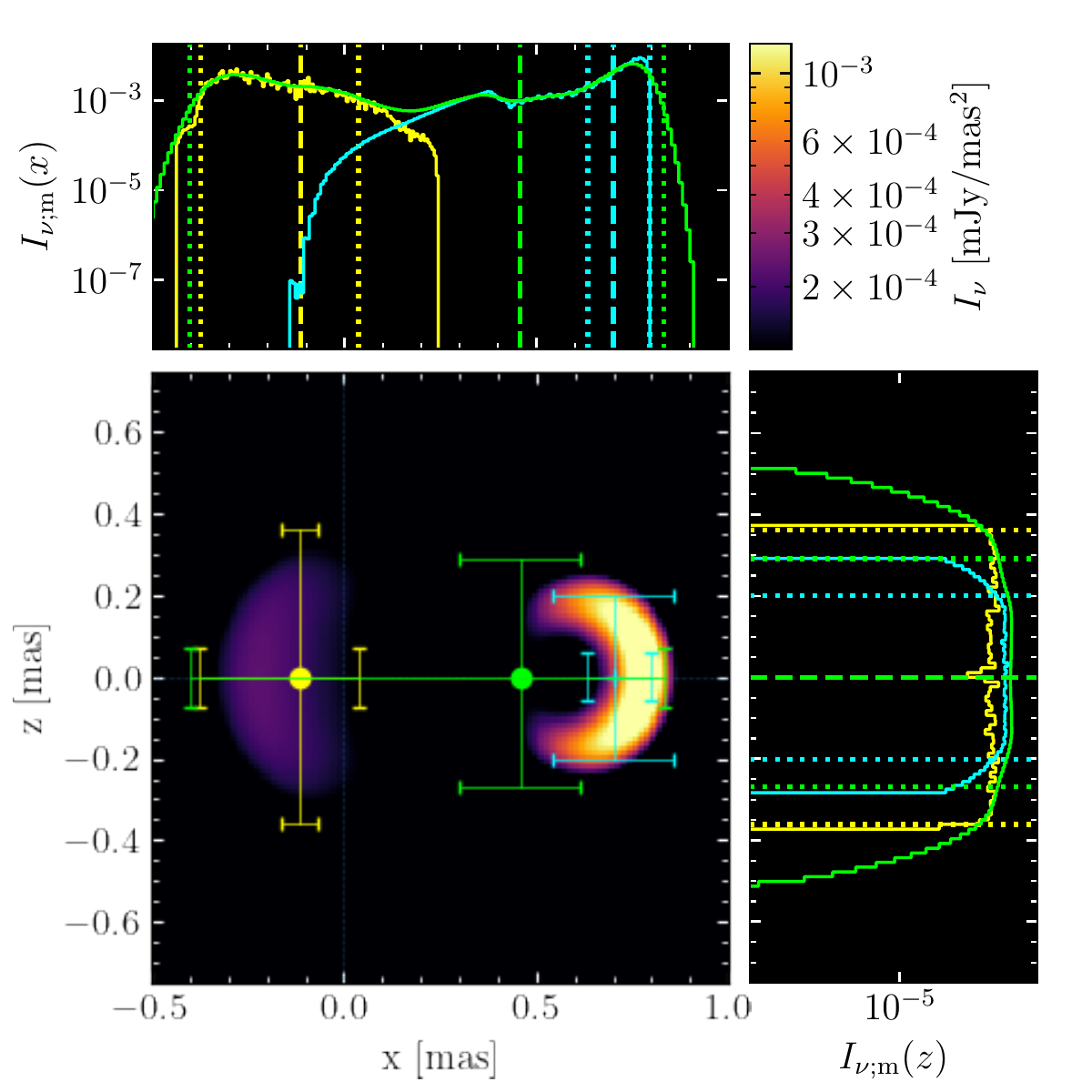}
    \caption{
        Combined \ac{kN} and \ac{GRB} sky map.
        The former can be seen as a dim blob on the left, while the 
        latter -- as a bright crescent on the right. 
        The size and the location of the flux centroid of two individual 
        components are shown with yellow and cyan colors respectively. 
        The size and $X_c$ of the combined image are shown as lime color. 
        As in Fig.~\ref{fig:results:skymap_example} the top and right 
        sub-panels display the $z$- and $x$-averaged brightness 
        distributions respectively. 
        Sky map corresponds to 
        $\nu_{\rm obs}=1\,$GHz, 
        $\theta_{\rm obs}=45\,$deg 
        and $t_{\rm obs}=60\,$days, 
        $n_{\rm ISM}=0.00031\,\ccm$.
    } 
    \label{fig:results:kn_grb_skymap_example}
\end{figure}

\begin{figure*}
    \centering 
    \includegraphics[width=0.49\textwidth]{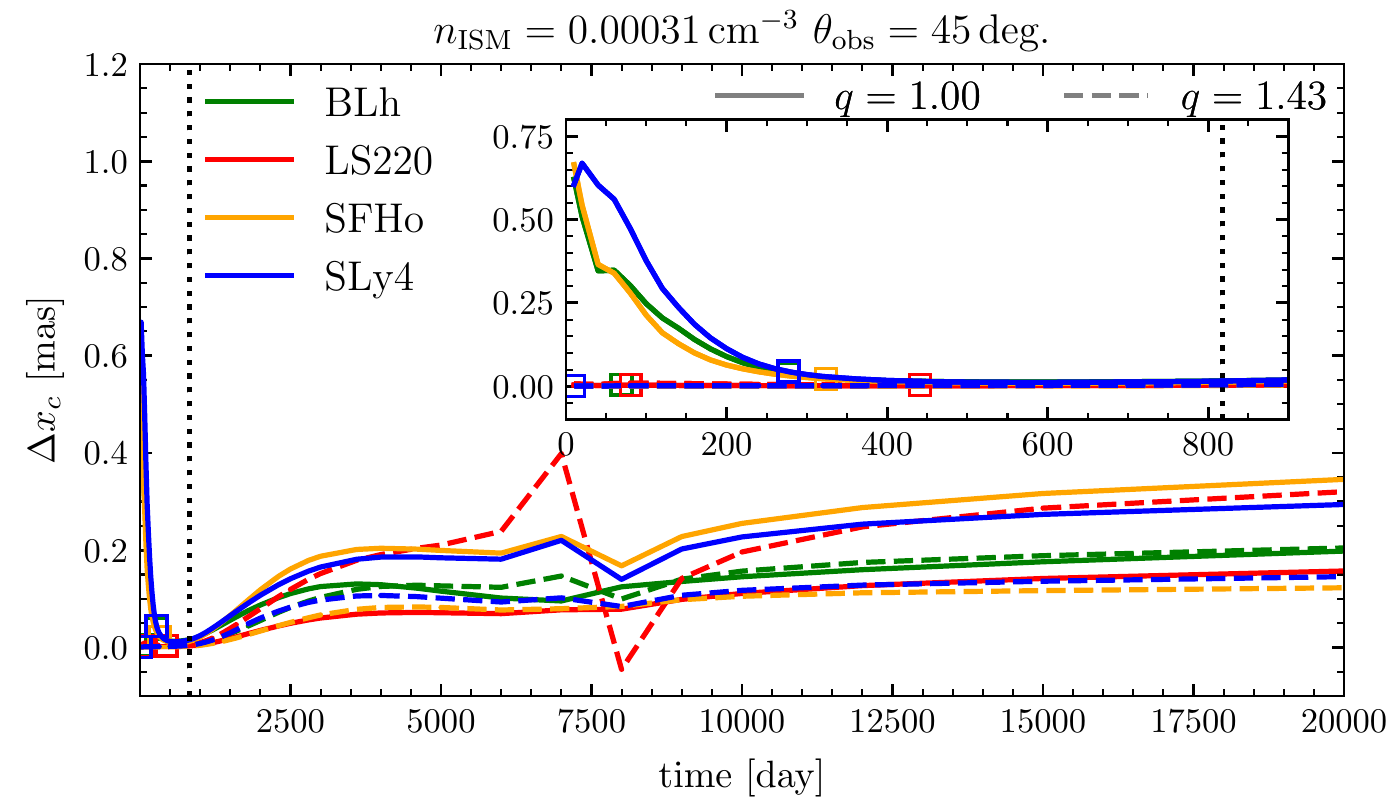}
    \includegraphics[width=0.49\textwidth]{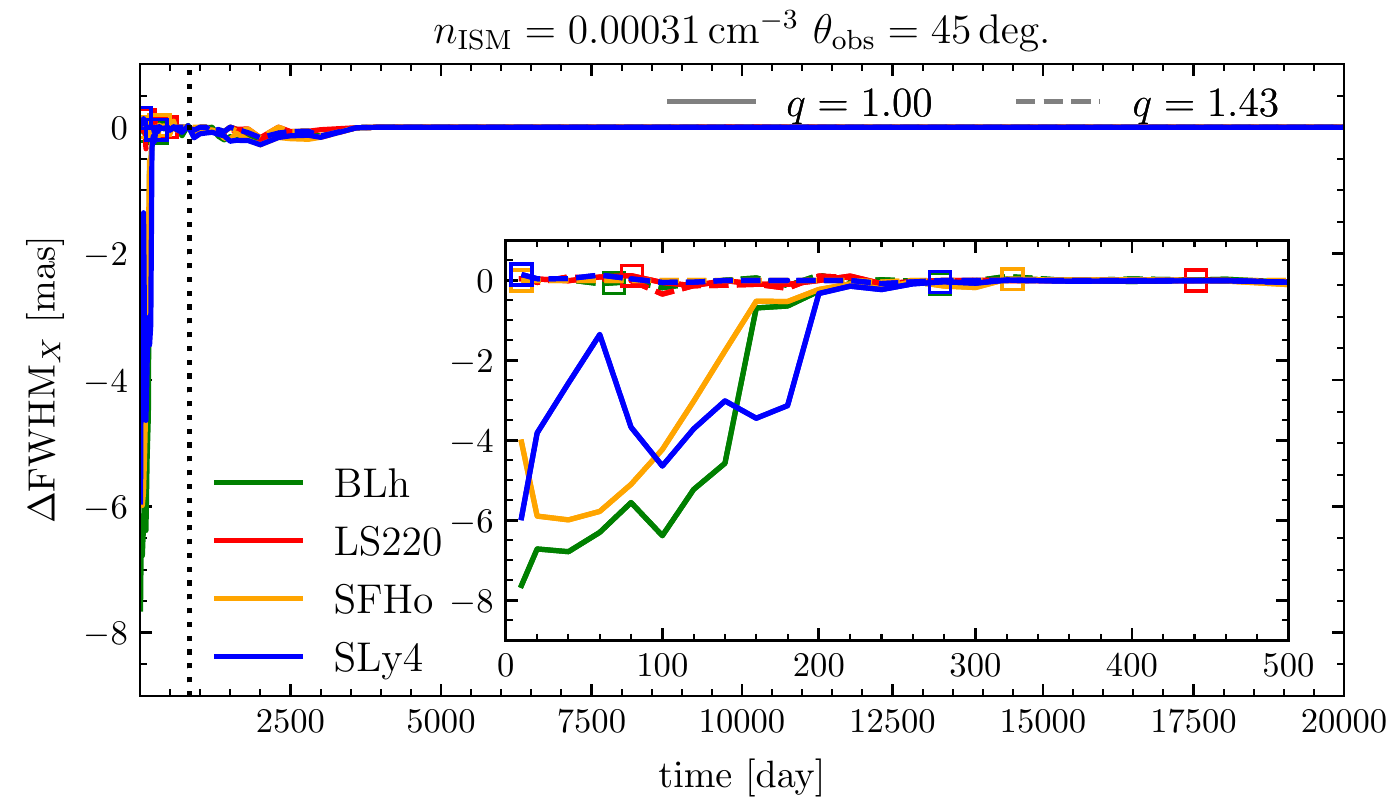}
    \includegraphics[width=0.49\textwidth]{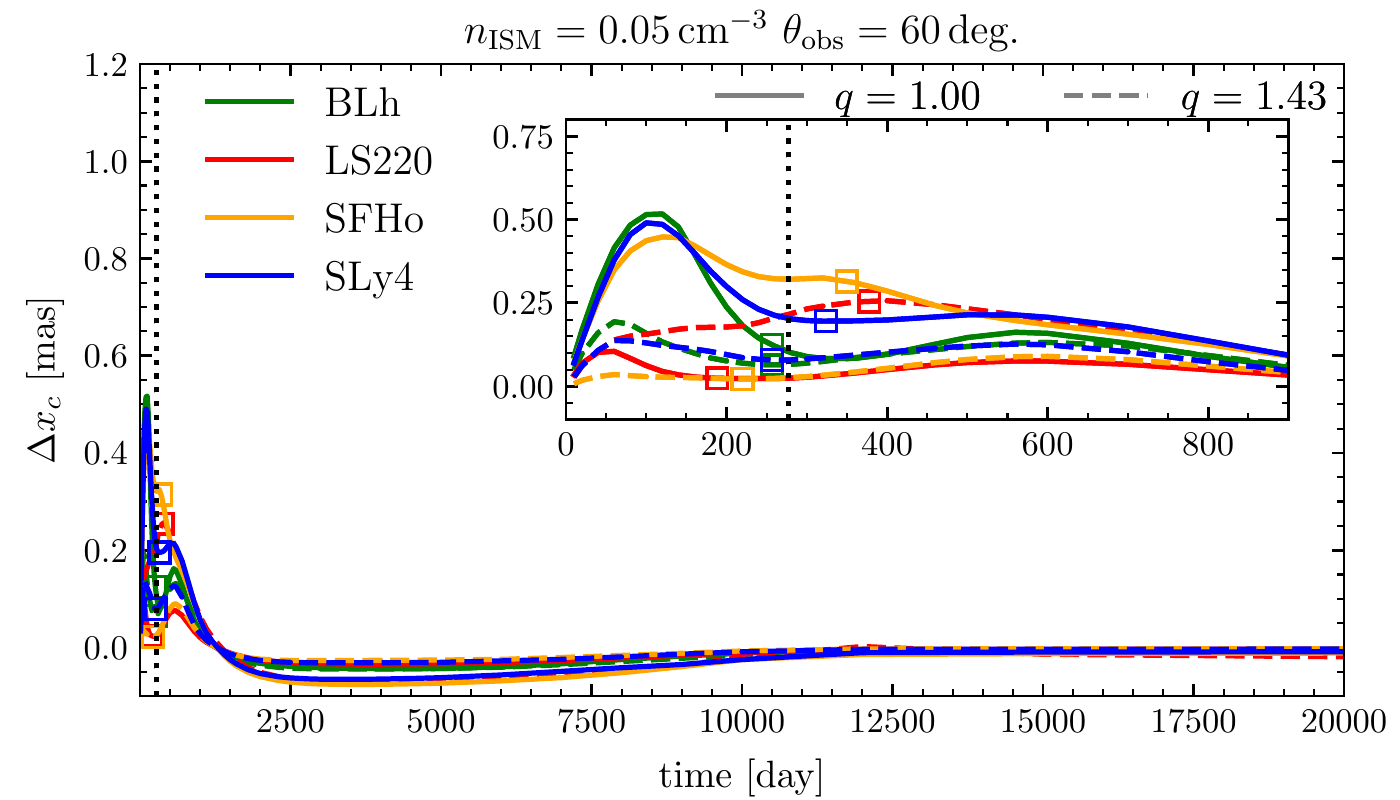}
    \includegraphics[width=0.49\textwidth]{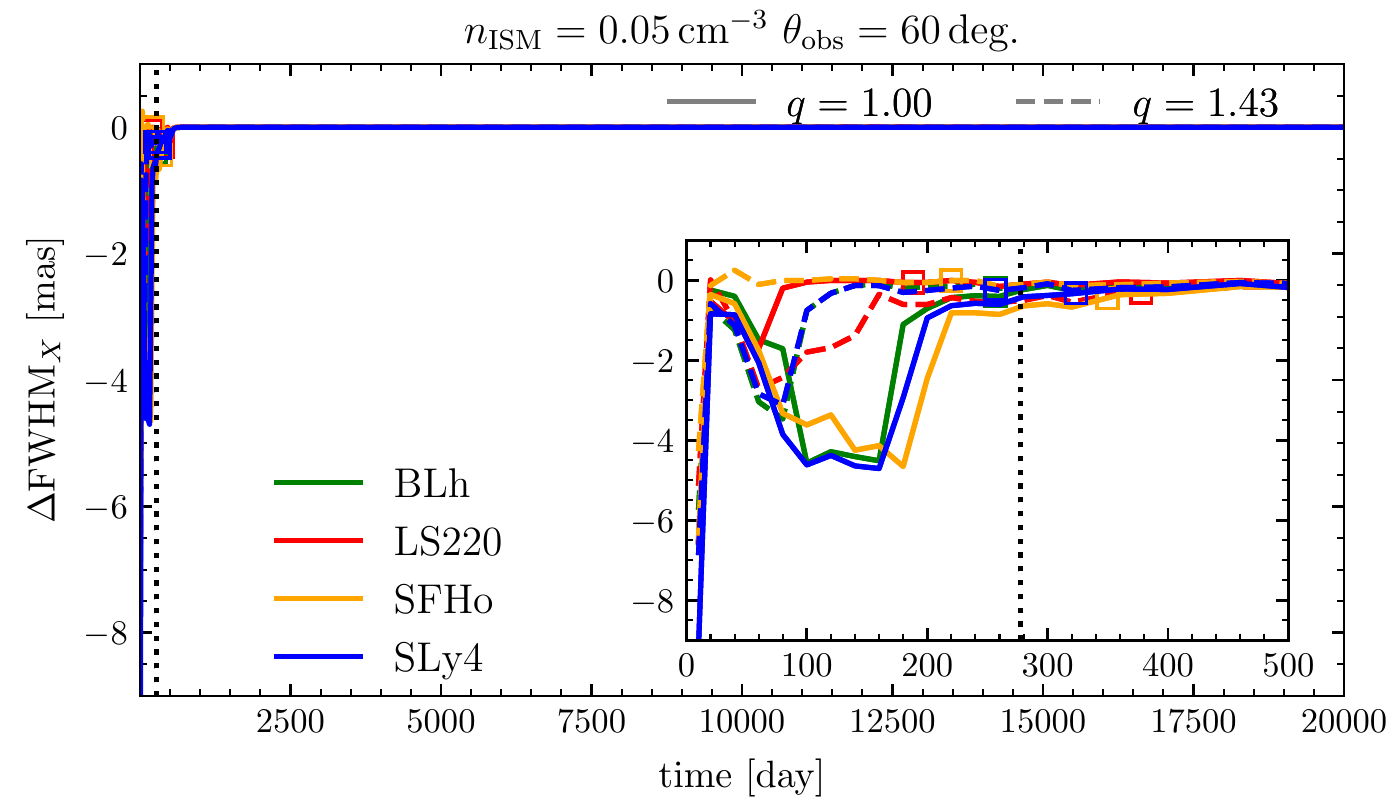}
    \includegraphics[width=0.49\textwidth]{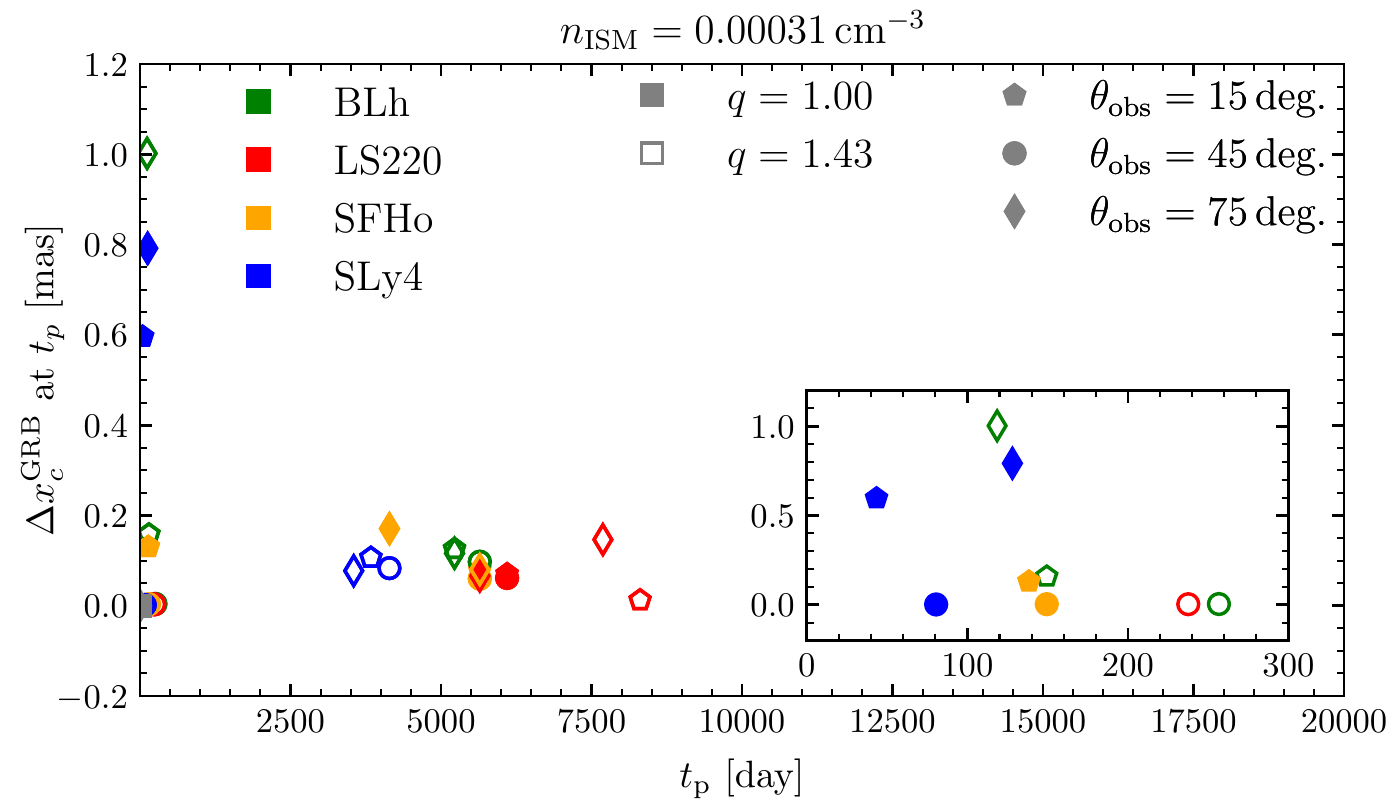}
    \includegraphics[width=0.49\textwidth]{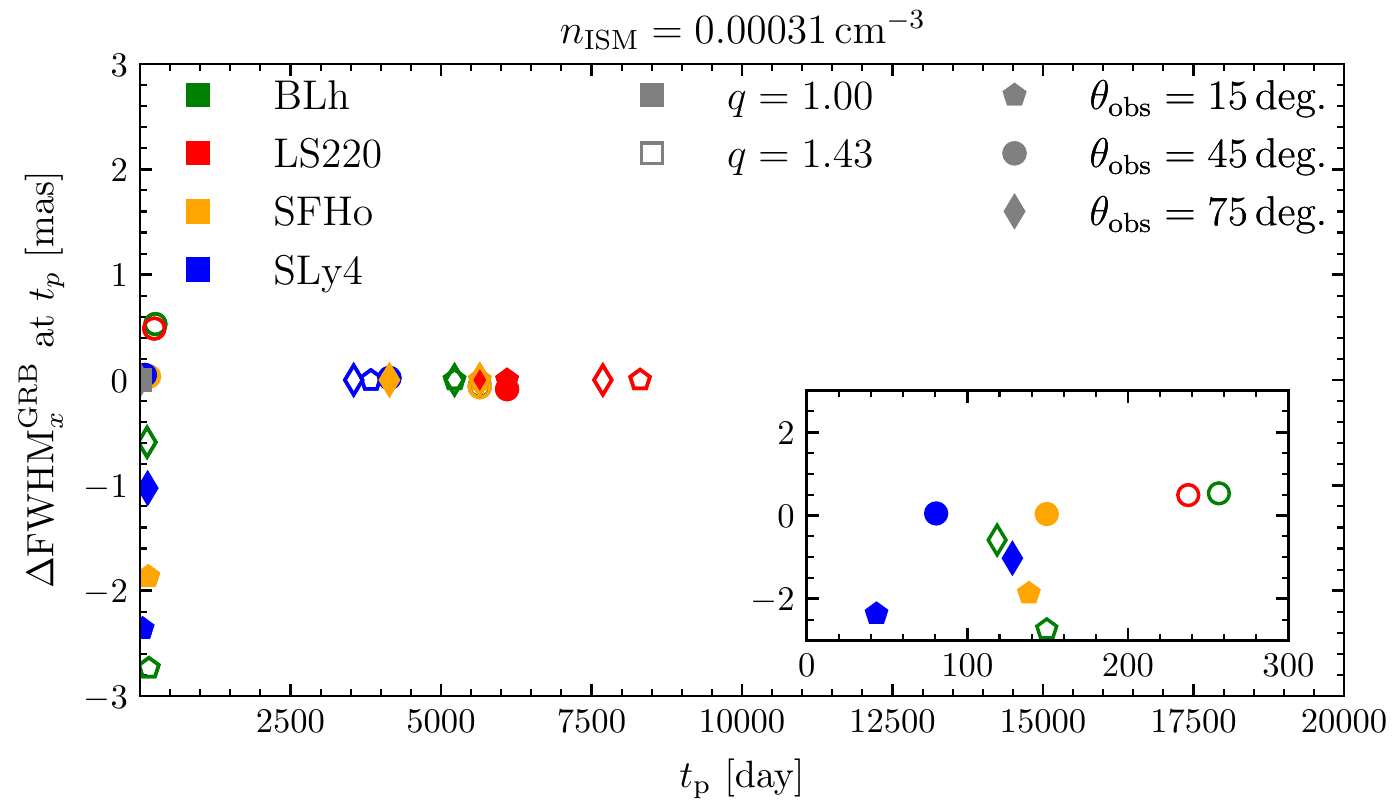}
    \caption{
        \textit{Top panels}: time evolution of the combined sky map properties 
        shown in terms of the 
        $\Delta x_c^{\rm GRB} = (x_c^{\rm GRB} - x_c^{\rm kN})/x_c^{\rm GRB}$, $\Delta \text{FWHM}_x^{\rm GRB} = \text{FWHM}_x^{\rm GRB} - \text{FWHM}_x^{\rm kN}$ on the \textit{left} and \textit{right} 
        panels respectively. 
        Dashed gray line corresponds to the time of the \ac{GRB} \ac{LC} 
        peak. 
        \textit{Bottom panels}: properties of the combined sky map 
        extracted at the time of the \ac{kN} afterglow \ac{LC} peak.
        Different colors correspond to various \acp{EOS}.
        Filled and empty markers indicate $q=1.00$ and $q=1.43$ simulations 
        respectively.
        Different markers correspond to various observing angles. 
        In all panels, an inner sub-panel serves to enlarge the early time 
        part of the figure. 
    } 
    \label{fig:results:kn_grb_skymaps_xc_evol_all_sims}
\end{figure*}

In Fig.~\ref{fig:results:kn_grb_skymap_example}, we show a combined 
\ac{kN} plus \ac{GRB} afterglow radio sky map assuming 
$\theta_{\rm obs}=45\,$deg. and $t_{\rm obs}=60\,$days. 
At this early time the \ac{GRB} afterglow is significantly brighter 
than the \ac{kN} one: 
$F_{\nu=1\,{\rm GHz}}^{\rm GRB}=7.5\times10^{-3}\,$mJy and 
$F_{\nu=1\,{\rm GHz}}^{\rm kN}=4\times10^{-4}\,$mJy. 
However, despite being dimmer, \ac{kN} afterglow affects the  
properties of the total sky map significantly, shifting the 
position of the image flux centroid back to the center of the explosion. 
Consequently, the apparent velocity computed from the motion 
of the flux centroid would be underestimated if the effect of \ac{kN} 
afterglow is not taken into account. In our case, the apparent 
velocity is reduced from $2.5\,c$ to $2.1\,c$ at $t_{\rm obs}=60\,$days. 
Thus systematic underestimation of the apparent velocity may, in turn, 
result in overestimation of the $\theta_{\rm obs}$ or $\Gamma$. 
This can be understood from the following considerations. Consider, 
$(\theta_{\rm s} \leq \theta_{\rm obs} - \theta_{\rm s})$, 
where $\theta_s$ is the average size of the extended source. 
There, the maximum apparent velocity $\beta_{\rm app}$ is equal 
to the source \ac{LF} $\Gamma$, as $\theta_{\rm obs} = 1/\Gamma$. 
Then, assuming that the observed emission from an extended source comes 
predominantly from the compact region we have, 
$(\theta_{\rm obs} - \theta_{\rm s}) \approxeq 1/\beta_{\rm app}$. 
These arguments were used to infer $\Gamma$ from radio image 
for \GRB{} \cite{Mooley:2018dlz}. 

Notably, at smaller observational angles, the early \ac{GRB} afterglow 
is significantly brighter, and at $\theta_{\rm obs}\simeq20\,$deg. 
that is generally inferred for \GRB{}, the \ac{kN} afterglow does 
not affect the estimated $\beta_{\rm app}$ to an appreciable degree.

At slightly later times, when the \ac{GRB} afterglow reaches its peak 
emission we find that even for $\theta_{\rm obs}=45\,$deg, the effect 
of the \ac{kN} afterglow on the \ac{GRB} afterglow sky map properties 
is negligible. At the time of the \ac{GRB} \ac{LC} peak 
$t_{\rm p}^{\rm GRB}=800\,$days, the $\beta_{\rm app}$ is 
reduced only by ${\simeq}0.1\,c$. 

The \ac{kN} afterglow becomes important again later, when the \ac{GRB} 
afterglow emission subsides. Numerical and semi-analytic jet models show, 
that both prime and counter jets contribute to the late time flux 
\cite{Zrake:2018eml,Fernandez:2021xce}. This forces the position of the flux 
centroid to move back to $x_c^{\rm GRB}\rightarrow 0$. Before that, the 
jet deceleration reduces the contribution to the observed emission from the 
fast jet core and consequently slows down the motion of the flux centroid. 
The jet lateral spreading contributes to this by pushing parts of the 
jet to $\theta > \theta_{\rm obs}$, making them move back on the image 
plane. 
In this regard, the presence of a \ac{kN} afterglow might be confused with 
a more rapid lateral spreading or earlier emergence of the counter jet.
\textit{
Thus, we conclude that even if the \ac{kN} afterglow does not contribute significantly to the observed total flux, it should be 
 taken into account for accurate estimation of the jet energy and 
geometry from sky map observations. 
}
Importantly, the relative brightness of two afterglows considered 
here depends on all free parameters of the model \ie, 
microphysics parameters of both shock types 
(relativistic and mildly relativistic) as well as the angular 
and velocity structure of ejecta.

Considering the available \ac{BNS} merger simulations, we recall 
that the \ac{kN} afterglow from $q=1$ and soft \acp{EOS} simulations 
is brighter, and thus it would affect the properties of the combined 
sky map more strongly, at least before the \ac{GRB} \ac{LC} 
peak $t_{\rm p}^{\rm GRB}$. 
In Fig.~\ref{fig:results:kn_grb_skymaps_xc_evol_all_sims}, 
we show the change in \ac{GRB} afterglow $x_c^{\rm GRB}$ and 
\size{} in terms of $\Delta v = (v^{\rm GRB}-v^{\rm kN}) / v^{\rm GRB}$, 
where $v\in[x_c,\text{\size{}}]$. 
As expected, the general effect of the inclusion of the \ac{kN}
afterglow is the decrease in $X_c$ and, consequently, in
the apparent velocity $\beta_{\rm app}$, and an increase in 
the image \size{} (top right and left panels of Fig.~\ref{fig:results:kn_grb_skymaps_xc_evol_all_sims}). 
Specifically, $\Delta x_c$ and $\Delta$\size{} reach 
${\gtrsim}0.5$ and ${\gtrsim}-8$ respectively.

At $t_{\rm p}^{\rm GRB}$ the effect of the \ac{kN} afterglow presence 
is minimal in all cases, as the \ac{GRB} afterglow dominates the total 
emission and the sky map properties. 
\textit{
Thus, estimated at this time, image properties convey the most reliable 
information about the \ac{GRB} afterglow. 
}

At higher $n_{\rm ISM}$ and $\theta_{\rm obs}$, the picture is 
qualitatively similar. Influence of the \ac{kN} afterglow is the most 
prominent at $t < t_{\rm p}^{\rm GRB}$ and for equal mass \ac{BNS} 
simulations with soft \ac{EOS}, such as SLy4 and SFHo \acp{EOS}. 
For $q>1$ simulations, the maximum $\Delta x_c$ and $\Delta$\size{} 
are about two times smaller than in $q=1$ cases. 
On the other hand, at $\theta_{\rm obs}=21.5\,$deg., and 
$n_{\rm ISM}=0.00031\,\ccm$, the influence of the \ac{kN} afterglow is 
negligible even at $t<t_{\rm p}^{\rm GRB}$ for all simulations. 
In this case \ac{GRB} afterglow provides a dominant contribution 
to the total \ac{LC} and the sky map, and the presence of \ac{kN} 
afterglow can only be seen at very late times 
$t \gg t_{\rm p}^{\rm GRB}$, when the \ac{kN} afterglow emission 
is coming predominantly from the non-thermal electron population. 
Meanwhile, in cases when the early \ac{GRB} emission is beamed away, 
$\theta_{\rm obs} \gtrsim 45\,$deg, the maximum in $\Delta x_c$ 
and $\Delta$\size{} occurs before the extreme in \ac{kN} afterglow 
spectral index evolution, in the regime where the emission from 
thermal electrons dominate the observed flux.

\subsection{Effect of the GRB-modified ISM on kN afterglow sky map}\label{sec:result:interaction}

\begin{figure}
    \centering 
    \includegraphics[width=0.49\textwidth]{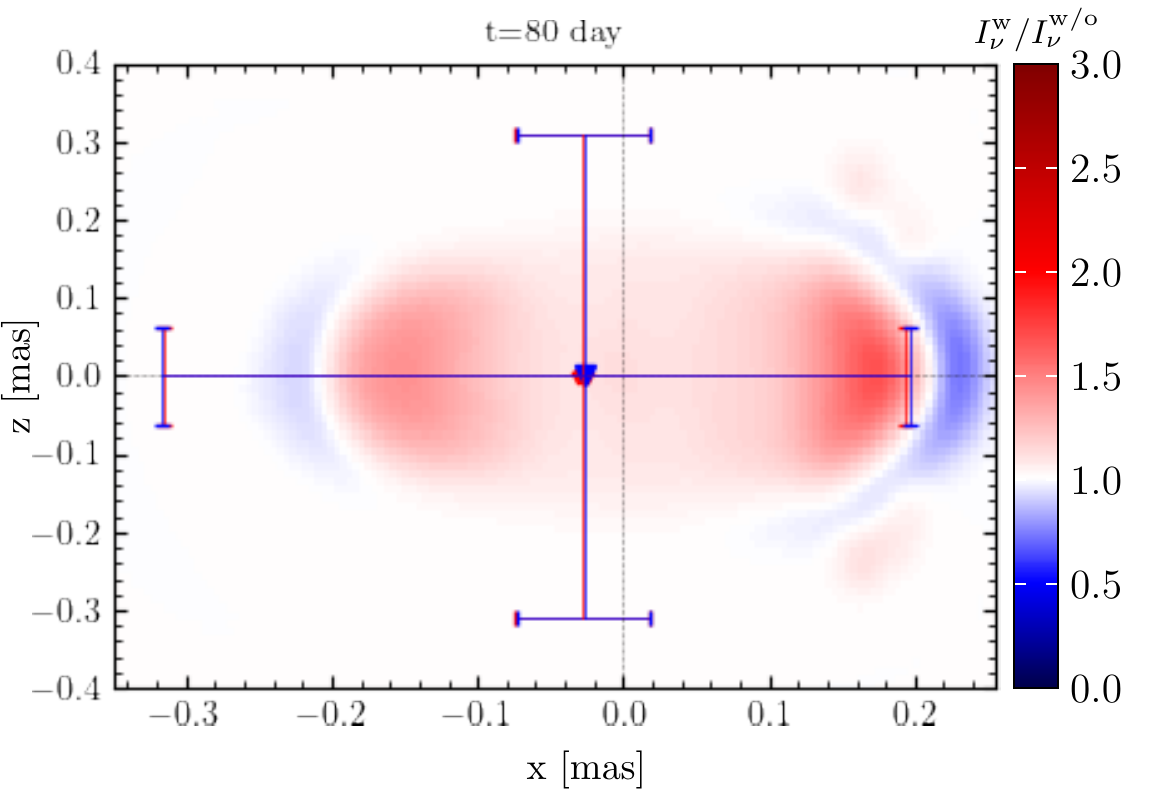}
    \caption{
        The ratio between two \ac{kN} afterglow sky maps with  
        the only difference between them being is whether 
        the \ac{CBM}, altered by a passage of \ac{GRB} \acp{BW},  
        is taken into account ($I_{\nu}^{\rm w}$) or not 
        ($I_{\nu}^{w/o}$). 
        Image size and the position of the flux centroid are 
        shown as before with error bars and markers with  
        blue color for ``w'' case and red for ``w/o'' case. 
        Sky maps are computed assuming $\nu_{\rm obs}=1\,$GHz, 
        $\theta_{\rm obs}=60\,$deg., and $n_{\rm ISM}=0.05\,\ccm$.
        } 
    \label{fig:results:kn_wg_example}
\end{figure}

In \citetalias{Nedora:2022kjv} we showed that when the \ac{kN} ejecta moves 
behind the \ac{GRB} \ac{BW}, it encounters an altered density profile, 
that we called an altered \ac{CBM}, and the afterglow signature changes. 
Specifically, 
the observed flux first decreases as most of the \ac{kN} ejecta moves 
subsonically behind the laterally spreading \ac{GRB} \ac{BW}, then 
increases as the \ac{kN} ejecta shocks the overdense fluid behind the 
\ac{GRB} \ac{BW} forward shock. However, the decrease and increase 
in the observed flux were found to be rather small: $\lesssim40\%$ and 
$\lesssim10\%$, respectively. The reason for this is the non-uniform 
nature of the \ac{kN} ejecta and finite time that \ac{GRB} lateral 
spreading takes. Thus, different parts of the \ac{kN} ejecta encounter 
different regions of the altered \ac{CBM} at a given time producing 
either an excess or a reduction in observed emission. 
Nevertheless, for the sake of completeness, it is worth looking at 
how the \ac{kN} afterglow sky map changes the altered \ac{CBM} is taken 
into account. 

In Fig.~\ref{fig:results:kn_wg_example} we show the effect of an 
altered \ac{CBM} on the \ac{kN} afterglow sky map for 
$t_{\rm obs}=80\,$days, $\theta_{\rm obs}=60\,$deg., and 
$n_{\rm ISM}=0.05\,\ccm$. The red and blue colors indicate the excess 
and the reduction of the observed emission with respect to the sky map 
computed when the altered \ac{CBM} is not taken into account. 
As expected, the change in the observed intensity occurs primarily 
near poles ($z=0$) and corresponds to \ac{kN} ejecta moving subsonically 
and not producing synchrotron emission. 
Fast elements of the \ac{kN} ejecta shocked the overdense region 
behind the \ac{GRB} shock and produced an emission excess. 
Slower elements of ejecta catch up with the underdense part of the 
altered \ac{CBM} later and this the part of the image where the emission 
is suppressed lies ahead of the one with emission excess. 
The more equatorial part of the ejecta avoids interacting with the 
altered \ac{CBM} and, thus, its emission remains unchanged (along $z$ axis). 
The certain parts of the image, the emission excess 
can be significant, ($I_{\nu}^{\rm w}/I_{\nu}^{\rm w/o}\lesssim3$). 
However, combined with the emission suppression in other parts of the 
image, the overall emission excess is rather small. 
Thus, even at this relatively high $n_{\rm ISM}$ and large 
$\theta_{\rm obs}$ the effect of the altered \ac{CBM} on the sky map 
properties, \ie, the position of the flux centroid and the image size are 
negligible.

\section{Discussion and conclusion}\label{sec:conclusion}

In this work, we considered synthetic radio images of the 
\ac{GRB} and \ac{kN} afterglow. 
For the former we considered \GRB{} motivated model settings, 
\ie, laterally structured jet observed off-axis 
\cite{Hajela:2019mjy,Fernandez:2021xce}. 
For the latter, we considered a set of ejecta profiles from 
\ac{NR} \ac{BNS} merger simulations targeted to \GW{}, 
\ie, with corresponding chirp mass. 
For all calculations, we use the semi-analytic afterglow code \pba{}, 
presented and discussed in 
\citetalias{Nedora:2021eoj} and \citetalias{Nedora:2022kjv}. 
The key aspect of the input physics is the inclusion of 
two electron populations behind the \ac{kN} \ac{BW} shocks, 
that follow power-law (non-thermal electrons) 
and Maxwellian (thermal electrons) distributions. 

The main limitation of our work is the semi-analytical nature 
of the model we employ. It remains to be investigated how 
\ac{GRB} and \ac{kN} afterglow sky maps computed with 
\ac{HD} numerical codes compare to ours. It is however 
numerically very challenging to perform such simulations on 
a temporal and spatial scales discussed in this work, as well as, 
to perform them for various possible choices of the model 
free parameters and \ac{kN} ejecta profiles. 

The aforementioned limitations notwithstanding, we find that the 
\ac{kN} afterglow sky map at early times resemble a wheel or 
a doughnut due to the emission from thermal electrons 
enhanced by relativistic 
effects, dominating the observed flux. At later times, the 
sky map is largely spherical with a remaining ring structure 
reflecting the 
$a)$ assumed axial symmetry, 
$b)$ initial ejecta velocity distribution. 
The image size evolves monotonically, albeit not smoothly, 
reaching ${\simeq10}\,$\ac{mas} at $3000\,$days and 
${\simeq}25\,$\ac{mas} at $20000\,$days.
If the \ac{kN} afterglow \ac{LC} at its peak is dominated by the 
emission from thermal electrons, the image size is smaller 
reaching ${\lesssim}5\,$\ac{mas}. 
Thus, the properties of the fast ejecta tail can be inferred 
from the sky map size and its evolution. 

Despite asymmetry in ejecta velocity distribution, however, 
the position of the image flux centroid $x_c$ does not deviate 
much from $0$, and is the largest ($|x_c| < 0.4\,$\ac{mas}) 
at early times, in cases when the emission from thermal electrons 
dominates the observed flux. Notably, however, the asymmetry can 
lead to the negative values of $|x_c|$ 
(assuming more on-axis observers), which if observed might hint 
at the equatorial nature of the fast ejecta tail.

Crucially, the presence of the \ac{kN} ejecta can affect 
the \ac{GRB} afterglow sky map to an appreciable degree 
even if the former does not appreciably contribute 
to the total observed flux. 
For that to occur, however, the source must be observed 
sufficiently off-axis so that the early \ac{GRB} afterglow 
emission is beamed away, while the \ac{kN} afterglow emission, 
dominated at this time by the emission from thermal electrons,  
is instead beamed more toward an observer. 
Specifically, at $t_{\rm obs}=80\,$days and assuming 
$\theta_{\rm obs}=45\,$deg. the change in the inferred value 
of the apparent velocity $\beta_{\rm app}$ can reach $0.5\,c$. 
At smaller $\theta_{\rm obs}$ the \ac{kN} afterglow effects 
the \ac{GRB} afterglow sky map properties significantly less and 
at $\theta_{\rm obs}\simeq20\,$deg. we find the effect to be 
negligible. 
Importantly, the relative brightness between these two types 
of afterglow depends on their respective sets of free parameters 
that are largely unconstrained. It is thus important to conduct 
a more thorough statistical analysis of the combined parameter 
space to assess the upper and lower limits of the degree to 
which the \ac{kN} afterglow influences the combined 
sky map properties.  

\def\ngVLA{Eddins et. al. (2023, in prep.)}

The detectability of the \ac{kN} and \ac{GRB} sky maps with 
Next Generation Very Large Array (ngVLA), which is currently in 
the development will be discussed in a separate study by \ngVLA{}. 
Overall, in order for \ac{kN} afterglow itself to be detectable, the 
flux density at the \ac{LC} peak should be $\gtrsim 5\times10^{-3}\,$mJy 
in radio \cite{Kathirgamaraju:2019xwu}. For \ac{BNS} merger 
simulations considered here, this is only possible at sufficiently 
high density, $n_{\rm ISM}\gtrsim0.005\,\ccm$ at $D_{\rm L}\simeq40\,$Mpc. In order to distinguish 
\ac{GRB} and \ac{kN} afterglows, the $\theta_{\rm obs}$ should be 
much larger than the jet opening angle (\eg, see figure~9 in 
\citetalias{Nedora:2022kjv}). 
At the same time, at large $\theta_{\rm obs}$ the change in the 
position of the sky map flux centroid due to the presence of the 
\ac{kN} afterglow can become detectable. It is, however, difficult to 
determine what value of $x_{c}^{\rm GRB}-x_{c}^{\rm kN}$ can be resolved.
At $n_{\rm ISM}=0.05\,\ccm$ and $\theta_{\rm obs}=60\,$deg, the 
$x_{c}^{\rm GRB}-x_{c}^{\rm kN}$ reaches $0.5\,$\ac{mas} for 
equal mass \ac{BNS} models within the first $200\,$days after 
the burst which, in principle, should be detectable (\ngVLA) 
with angular resolution of $0.1\,$\ac{mas}.

\section*{Acknowledgements}

The simulations were performed on the national supercomputer HPE Apollo Hawk at the High Performance Computing (HPC) Center Stuttgart (HLRS) under the grant number GWanalysis/44189 and on the GCS Supercomputer SuperMUC at Leibniz Supercomputing Centre (LRZ) [project pn29ba].

\textit{Software:} We are grateful to the countless developers 
contributing to open source projects that was used in the analysis 
of the simulation results of this 
work: \texttt{NumPy} \citep{numpy}, \texttt{Matplotlib} \cite{matplotlib}, and \texttt{SciPy} \cite{scipy}.

\section*{Data Availability:} 
The datasets generated during and/or analyzed during the current 
study are available from the corresponding author on reasonable request.

\appendix

\bibliography{refs20230224}

\end{document}